\documentclass[a4paper,11pt]{article}
\usepackage{jheppub} 
\usepackage[T1]{fontenc} 
\usepackage{bm} 
\usepackage{mathrsfs} 
\usepackage{subcaption}
\usepackage{mdframed}
\usepackage{enumitem}
\usepackage{multicol}
\usepackage{xcolor}
\usepackage{subcaption}
\usepackage{mathtools}
\usepackage{tikz-cd}
\usepackage[normalem]{ulem}
\usepackage{amsmath}
\usepackage{amssymb}
\usepackage{amsfonts}
\usepackage{latexsym}
\newmdenv[skipabove=6mm]{kotak}

\newcommand{\be}{\begin{align}}
\newcommand{\ee}{\end{align}}

\newcommand{\dif}{{\rm d}}

\newcommand{\downIn}{\rotatebox[origin=c]{90}{$\in$}}
\newcommand{\FuncSpace}[1]{\mathcal{F}(#1)}
\newcommand{\PrEoS}{\mathcal{P}_\mathrm{EoS}}
\newcommand{\PrObs}{\mathcal{P}_\mathrm{obs}}
\newcommand{\MapInv}{\Psi_{\mathrm{inv}}}
\newcommand{\MapTOV}{\Psi_\mathrm{TOV}}

\newcommand{\pc}{p_{\mathrm{c}}}
\newcommand{\NSNumber}{n_{\mathrm{NS}}}
\newcommand{\UniDis}{U}

\preprint{INT-PUB-24-001, YITP-24-05}

\title{\boldmath Uncertainty quantification in the machine-learning inference from neutron star probability distribution to the equation of state}

\date{}

\affiliation[a]{Institute for Nuclear Theory, University of Washington, Box 351550, Seattle, WA 98195, USA}

\affiliation[b]{Department of Physics, The University of Tokyo,
  7-3-1 Hongo, Bunkyo-ku, Tokyo 113-0033, Japan}
\affiliation[c]{Yukawa Institute for Theoretical Physics, Kyoto University, Kyoto 606-8502 Japan}
\affiliation[d]{Department of Physics, Tokyo Metropolitan University, Hachioji 192-0397, Japan}

\abstract{
    We discuss the machine-learning inference and uncertainty quantification for the equation of state (EoS) of the neutron star (NS) matter directly using the NS probability distribution from the observations.
    We previously proposed a prescription for uncertainty quantification based on ensemble learning by evaluating output variance from independently trained models.  We adopt a different principle for uncertainty quantification to confirm the reliability of our previous results.  To this end, we carry out the MC sampling of data to infer an EoS and take the convolution with the probability distribution of the observational data.  In this newly proposed method, we can deal with arbitrary probability distribution not relying on the Gaussian approximation.  
    We incorporate observational data from the recent multimessenger sources including precise mass measurements and radius measurements.
    We also quantify the importance of data augmentation and the effects of prior dependence.
}

\date{}

\author[a]{Yuki Fujimoto}
\author[b]{Kenji Fukushima}
\author[b]{Syo Kamata}
\author[c,d]{Koichi Murase}

\emailAdd{yfuji@uw.edu}
\emailAdd{fuku@nt.phys.s.u-tokyo.ac.jp}
\emailAdd{skamata11phys@gmail.com}
\emailAdd{phys.murase@gmail.com}

\begin{document} 
\maketitle
\flushbottom

\section{Introduction}
\label{sec:introduction}

Unveiling properties of QCD (Quantum Chromodynamics) matter in the regime of finite baryon density is a vital challenge in modern nuclear physics.
Despite various efforts over several decades, many aspects of finite-density QCD have remained elusive, such as the nature of QCD transitions (see refs.~\cite{Fukushima:2010bq, Baym:2017whm, Aarts:2023vsf} for comprehensive reviews).
The equation of state (EoS) of dense matter is of central importance in describing the properties of QCD matter.
Ample knowledge of the EoS, that is, the behavior of thermodynamic quantities in extreme environments, would help us reveal the properties of states emerging from QCD\@.
This approach has indeed worked successfully in the case of QCD at small density and high temperature, where the EoS from numerical simulations of first-principles lattice QCD is the bedrock for the crossover from the hadronic state to a quark-gluon plasma~\cite{Borsanyi:2013bia, HotQCD:2014kol, HotQCD:2018pds}.
By contrast, the finite-density state is inaccessible by the Monte-Carlo (MC) algorithm due to the sign problem, though there are a few useful exceptions and one can perform the finite-density simulations in lattice QCD~\cite{Alford:1998sd, Son:2000xc, Iida:2022hyy, Brandt:2022hwy, Abbott:2023coj, Fujimoto:2023unl}.

Neutron stars (NSs) by far provide the most reliable and robust constraints on the dense-matter EoS in the region, $n_B \gtrsim 2 n_0$, where $n_B$ is the baryon density and $n_0 \simeq 0.16~\text{fm}^{-3}$ is the nuclear saturation density.
The NS observables include relationships between the mass $M$, the radius $R$, the tidal deformability $\Lambda$, etc., while the EoS is expressed as a relation between the pressure, $p$, and the energy density, $\varepsilon$.
The one-to-one correspondence between the NS observables and the EoS via the general relativistic equation called the TOV (Tolman-Oppenheimer-Volkoff) equation~\cite{Tolman:1939jz, Oppenheimer:1939ne, 1992ApJ...398..569L} can constrain possible EoS candidates.
Several studies have argued that it is feasible to anchor the EoS in the regions from $n_B \simeq n_0$ up to $\sim 2 n_0$ by microscopic calculations based on the chiral effective field theory~\cite{Hebeler:2010jx, Hebeler:2013nza, Gandolfi:2019zpj, Drischler:2020hwi, Drischler:2020fvz, Drischler:2021bup}.
One could also fix the very high-density part of the EoS around $n_B \gtrsim 40 n_0$ by the perturbative QCD (pQCD) calculations~\cite{Freedman:1976ub, Baluni:1977ms, Vuorinen:2003fs, Kurkela:2009gj, Gorda:2018gpy, Gorda:2021znl, Gorda:2021kme}.
With these setups, from the current lower/upper bounds for the heaviest known $M$'s~\cite{Demorest:2010bx, Antoniadis:2013pzd, Fonseca:2016tux, NANOGrav:2017wvv, NANOGrav:2019jur, Fonseca:2021wxt}, $R$'s~\cite{Riley:2019yda, Miller:2019cac, Miller:2021qha, Riley:2021pdl}, and $\Lambda$ in the merging binary NSs~\cite{LIGOScientific:2017vwq, LIGOScientific:2018cki, LIGOScientific:2018hze}, a large portion of the EoS has been strictly excluded~\cite{Kurkela:2014vha, Annala:2017llu, Annala:2019puf, Drischler:2020fvz, Annala:2021gom, Drischler:2021bup}.

To take one step further to obtain a likely candidate, one can perform statistical inference of the EoS from the observed NS data incorporating the uncertainty of astronomical observations by using various approaches.
The standard framework used in the community is Bayesian inference with a parametrized EoS developed in refs.~\cite{Steiner:2010fz, Steiner:2012xt, Ozel:2015fia}.
In a similar but more direct method discussed in earlier papers~\cite{Ozel:2009da, Ozel:2010fw}, the probability distribution of the observables is directly transformed into the probability of the EoS by the MC integration with a Jacobian marginalizing over the observables.
These Bayesian approaches with a parametrized EoS have been extended to a non-parametric formulation employing the Gaussian process, as introduced in ref.~\cite{Landry:2018prl}.
Machine learning (ML) using a deep neural network (NN) is another independent stream of research initiated in the series of works~\cite{Fujimoto:2017cdo, Fujimoto:2019hxv, Fujimoto:2021zas} by three of the present authors.
The problem of the EoS inference is attracting a growing interest in the nuclear astrophysics community, and different methods are beneficial for improving systematics.
Indeed there are many papers published even within the past few years along the lines of these approaches; an incomplete list of such studies includes refs.~\cite{Alvarez-Castillo:2016oln, Raaijmakers:2019dks, Greif:2020pju, Al-Mamun:2020vzu, Huth:2021bsp, Altiparmak:2022bke, Lim:2022fap, Chimanski:2022wzi, Marczenko:2022jhl, Brandes:2022nxa, Ecker:2022dlg, Jiang:2022tps, Takatsy:2023xzf, Annala:2023cwx, Brandes:2023hma, Musolino:2023edi, Richter:2023zec, Providencia:2023rxc, Cai:2023pkt, Pang:2023dqj,  Fan:2023spm, Tsang:2023vhh, Tang:2023owf, Komoltsev:2023zor, Imam:2024gfh} for Bayesian inference, refs.~\cite{Essick:2019ldf, Landry:2020vaw, Essick:2020flb, Essick:2020ghc, Essick:2021kjb, Legred:2021hdx, Essick:2021ezp, Legred:2022pyp, Gorda:2022jvk, Annala:2023cwx, Essick:2023fso, Mroczek:2023zxo, Komoltsev:2023zor} for Gaussian process, and refs.~\cite{Ferreira:2019bny, Morawski:2020izm, Traversi:2020dho, Krastev:2021reh, Soma:2022qnv, Thete:2022eif, Farrell:2022lfd, Soma:2022vbb, Ferreira:2022nwh, Goncalves:2022smd, Chatterjee:2023ecc, Krastev:2023fnh, Farrell:2023ojk, Zhou:2023cfs, Carvalho:2023ele, Carvalho:2024kgf} for machine learning.

The results in most literature suggest that the EoS exhibits a rapid stiffening at a few times $n_0$, i.e.\ a sharp increase in $p$ with increasing $\varepsilon$.
The stiffness of the EoS is characterized by the squared sound speed defined by the slope of the EoS, i.e., $c_s^2 := \dif p / \dif\varepsilon$, and it approaches $c_s^2 \to 1/3$ in the conformal limit at asymptotically high density.
At a few times $n_0$, $c_s^2$ may exceed $1/3$ such that the EoS is stiff enough to account for the existence of the heavy pulsars~\cite{Bedaque:2014sqa, Alford:2015dpa, Tews:2018kmu}.
It should be noted, however, that $c_s^2>1/3$ can be consistent with conformality that should be better characterized by the vanishing conformal anomaly $\varepsilon - 3p$ as $c_s^2$ is a derivative quantity~\cite{Fujimoto:2022ohj}.
Some of the literature claims that this rapid stiffening is followed by softening, resulting in a peak structure in $c_s^2$ as a function of $\varepsilon$~\cite{Fujimoto:2019hxv, Fujimoto:2021zas, Altiparmak:2022bke, Gorda:2022jvk}.
Anchoring the pQCD results at $n_B \simeq 40 n_0$ will help reduce $c_s^2$ at far higher densities out of the range of NS matter; whether pQCD significantly impacts the EoS in the NS density range is still under intense debates~\cite{Komoltsev:2021jzg, Somasundaram:2022ztm, Gorda:2022jvk, Brandes:2023hma, Mroczek:2023zxo, Komoltsev:2023zor}.
Such a peak structure in $c_s^2$ is generally expected if the EoS is bridged between the hadronic and the quark regions without a sharp phase transition~\cite{Masuda:2012ed, Masuda:2012kf, Kojo:2014rca, McLerran:2007qj, McLerran:2018hbz, Jeong:2019lhv, Kojo:2021ugu, Fujimoto:2022xhv, Fujimoto:2023ioi}, while purely nucleonic descriptions with the repulsive three-body forces~\cite{Akmal:1998cf, Hebeler:2009iv, Gandolfi:2011xu} or the vector mean field~\cite{Serot:1997xg} tend to increase stiffness monotonically with the density.
Thus, a peak structure in $c_s^2$ may signal the onset of quark degrees of freedom in the hadron-quark crossover picture.
If the peak structure is confirmed from the NS data without any theoretical bias, it strongly signifies the nontrivial change in underlying physics (see also ref.~\cite{Tan:2021nat} for possible detection through the binary Love relation).

Previously, we reported a rapidly stiffening EoS using the deep ML inference~\cite{Fujimoto:2019hxv, Fujimoto:2021zas}.
It should be noted that in what follows we use ``machine learning'' or ML to refer to our method specifically in terms of the NNs, as explained later, though Bayesian analysis can also be regarded as an ML technology in a general sense.
The sound speed from our previous ML method rises to $c_s^2 \simeq 0.61 \pm 0.16$ (with 68\% credence) around $\varepsilon / \varepsilon_0 \simeq 4$--$5$, where $\varepsilon_0$ is the energy density at $n_B = n_0$.  The high-density part around $\varepsilon/\varepsilon_0 \simeq 6$--$8$ reads $c_s^2 \simeq 0.55 \pm 0.10$, i.e., $c_s^2$ was slightly decreased as in Fig.~4 of ref.~\cite{Fujimoto:2019hxv}. 
One may thus identify a peak structure in $c_s^2$ as a function of $\varepsilon$, although the statistical significance was not enough.
Going into more detail, we first developed the method in ref.~\cite{Fujimoto:2017cdo} and extended it in ref.~\cite{Fujimoto:2019hxv};
we trained an NN with a sufficiently large dataset of mock observations increased by a few orders of magnitude with data augmentation.
For a given input of $(M,R)$ pairs of the NSs and their standard deviations, the NN outputs a set of parameters of the EoS, for which we employ a piecewise $c_s^2$.
Since there was no established method of uncertainty quantification for deep ML regression, the treatment of the input and the output uncertainties in our previous analysis assumed ad~hoc simplification.
Regarding the input uncertainties in the NS data, we extracted the standard deviations of the probability distributions of $(M,R)$ along the $M$ and $R$ directions by marginalizing the distributions.  In effect, we made the Gaussian approximation and used the standard deviations as the uncertainty quantifying parameters in the NN input.
For the output uncertainties, we attempted two different methods.
One is based on the comparison between the validation data for which we know the ground truth, and the other is based on the ensemble learning method, i.e., the method called bagging specifically.
We also recognized an issue with the high-density part of the inferred EoS being not well-constrained by the data at the time of refs.~\cite{Fujimoto:2019hxv, Fujimoto:2021zas}.

The purpose of this paper is to verify the nontrivial behavior of $c_s^2$ in our previous work by using an alternative method based on the convoluted analysis of the ML reconstruction of the EoS and the MC integration over the NN input space.
The present analysis can be regarded as complementary to our previous ones, and on top of that, the novel facets are found as follows:
(a) the MC integration handles the full shape of the probability distribution including the input uncertainty, and accordingly, it quantifies the output uncertainty,
(b) we take into account the new NS data including the heavier ones than those used in our previous studies, and
(c) we numerically check the effects of our prior assumptions on the inferred results.
In the following paragraphs, we will give further background to the above three points in order and motivate the present analysis.  We shall elaborate the technical details in the following sections.

First, we make full use of the entire information of the NS data, instead of using only the standard deviations of the marginalized distributions.
For this, we incorporate the original shape of the joint probability distribution of $M$ and $R$ into the analysis by the MC integration.
Each $(M,R)$ is sampled for the NN input and thus weighted by the probability;  through the MC integration over $(M,R)$-space, the uncertainties of the input data flow to the output data in the NN, that is, the EoS parameters in our setup.
In this way, our method proposed in this work is an alternative formulation of the uncertainty quantification for the ML inference.
In particular, we shall numerically confirm whether the input uncertainty propagates to the output uncertainty as it should in the proper uncertainty quantification.
We note that this way of uncertainty quantification is somehow similar to that in refs.~\cite{Ozel:2009da, Ozel:2010fw}, which additionally requires the explicit evaluation of the Jacobian of the variable transformation from $(M,R)$ to the EoS\@.
Treating the original shape of the probability distribution properly is demanded also by the fact that the joint probability distribution of $M$ and $R$ is significantly skewed.
In some astronomical measurements, moreover, the correlation between $M$ and $R$ is essential, e.g.\ the data from NICER (Neutron star Interior Composition Explorer) has a positive $R$-$M$ correlation because of the measurement of the compactness $M/R$.
In our previous prescription, we discarded the information on the $R$-$M$ correlation, as we dropped the finite skewness of the probability distributions.
These points comprise the main results of this work.

Secondly, as a straightforward extension, we bring our NS data up to date.
A major update since the last work is the radius measurement of heavy mass neutron star J0740+6620 from NICER~\cite{Miller:2021qha, Riley:2021pdl}.
Integrating the heavy NS data significantly enriches the information content in the high-density part of the EoS, which was lacking in our previous works.
We also take the heavy mass pulsars explicitly into account to inform the NN of the existence of such objects and incorporate the $M$-$R$ probability distribution of binary NSs in the GW170817 event which is inferred from the EoS-insensitive relation~\cite{LIGOScientific:2018hze}.
The data we used in ref.~\cite{Fujimoto:2019hxv} is the $M$-$R$ relation of NSs in photospheric radius expansion X-ray bursts (which we abbreviate as the PREs) and in the low-mass X-ray binaries in quiescent period (which we abbreviate as the qLMXBs)~\cite{Ozel:2015fia, Bogdanov:2016nle}.
These sets of data are obtained from \emph{spectral} measurements of X-rays and are known to contain a systematic uncertainty associated with the atmospheric composition of the star (see e.g.\ refs.~\cite{Miller:2013tca, Miller:2016pom} for a review), while the NICER performs \emph{timing} measurements of X-rays, so it is thought to be free from such systematic errors.
Thus, including a different source of data is in general beneficial in the viewpoint of controlling the systematic errors in the predicted EoS\@.

Finally, as our extended analysis, we shall examine and intentionally adjust several prior assumptions in the training dataset generation that may potentially affect the inferred results.
In our previous paper~\cite{Fujimoto:2021zas}, we have found that the inferred results in the density range unconstrained by the input data depend on the prior values;  it is partly responsible for decreasing behavior in $c_s^2$ at high densities, so it is crucial to check the dependence of the results on the prior values in our new analysis here.
In particular, we will consider three major factors:  (a) the EoS parametrization, (b) the $M$-$R$ data sampling in the input layer in the training dataset, and (c) the data augmentation.
We will elaborate procedures in secs.~\ref{sec:EoS} and \ref{sec:method.dataset}, and we will discuss related issues in section~\ref{sec:NN_property}.

The paper is organized as follows.
In section~\ref{sec:outline}, we review our previous method and its subtleties, and then conceptually introduce our new approach.
In section~\ref{sec:map}, we elaborate on the technicalities of the new method of mapping the $M$-$R$ data to the EoS using the NN and the MC integration.
Section~\ref{sec:results} presents our main results.
In section~\ref{sec:NN_property}, we perform numerical assessments for the validity of our results including the prior dependence checks in our NN model.
Throughout the paper, we use the natural unit: $c = G = 1$.

\section{Preview of the Present Work}
\label{sec:outline}

We shall overview our idea in advance to technical details about concrete analyses and results.  For the scientific application, it is crucial how to treat the uncertainties correctly for the input and output data.
In physics, any experimentally observed data carries an uncertainty in principle, which is usually estimated by the standard error $\sigma$, namely the standard deviation of the error.
For the inference problem from the input of the observational data to the output of the theoretical parameters, as we address in this work, we need to assess the standard errors in the output in response to the given uncertainty in the input.
In this section, we will revisit the previously adopted method and propose a new approach for the error estimate.

\subsection{Previous method}

\begin{figure}
    \centering
    \includegraphics[width=0.7\textwidth]{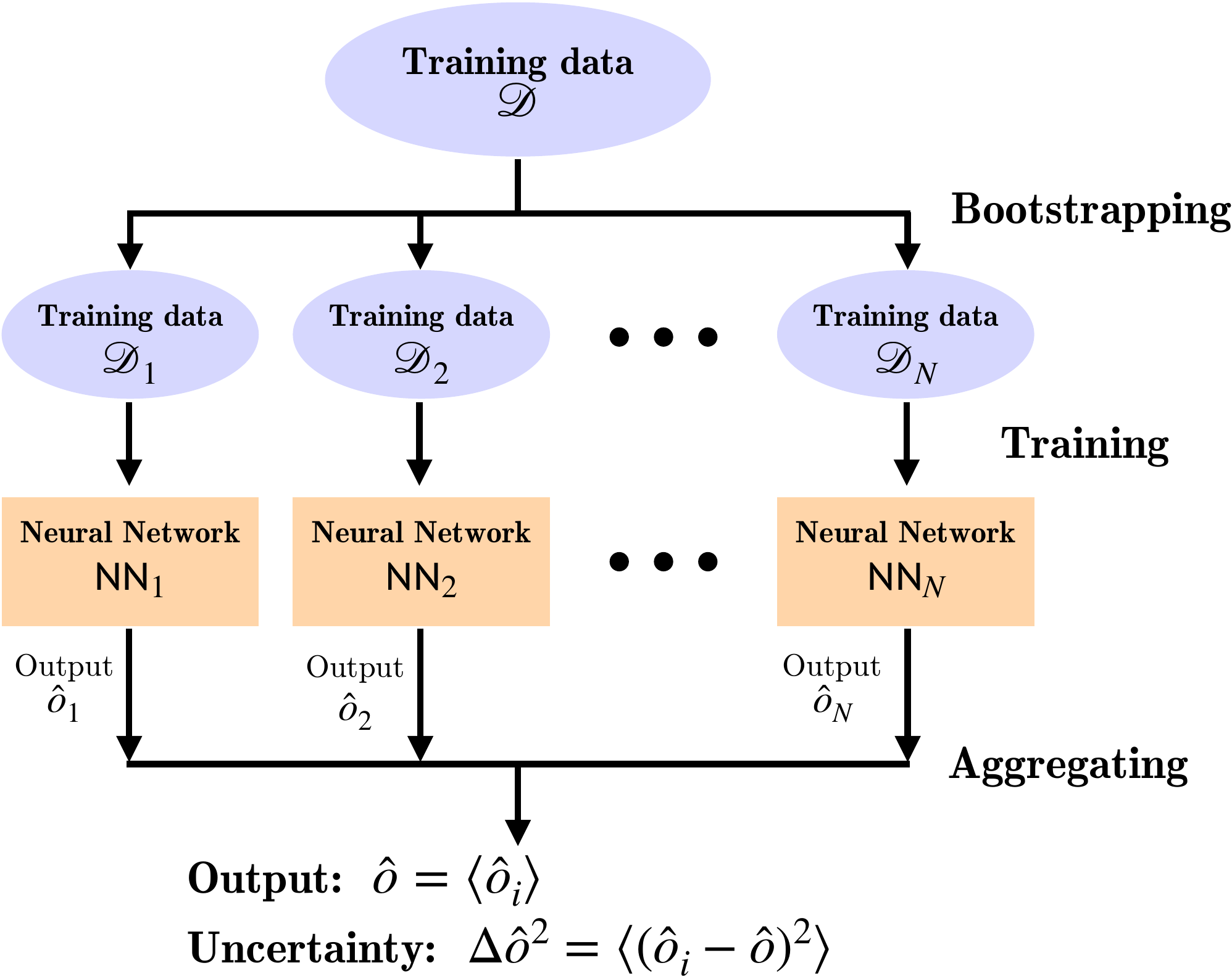}
    \caption{Previously adopted method: Bagging to estimate the uncertainty using the fluctuation from the parallel models.}
    \label{fig:bagging}
\end{figure}

In our previous works~\cite{Fujimoto:2017cdo,Fujimoto:2019hxv,Fujimoto:2021zas}, we proposed a method for the error estimate based on the bagging technique.
In this technique randomly chosen subsets of the training data are used to build parallel models for the purpose of independent training processes.
It should be noted that we use the term, bagging, in a loose sense as remarked in ref.~\cite{Fujimoto:2021zas}.
Then, the final output is given as a function (e.g.,\ mean) of all the outputs from the parallel models.
This bagging technique, often referred to as bootstrap aggregating (see figure~\ref{fig:bagging}), is useful particularly for coping with noisy data.  At the same time, we can characterize fluctuations associated with the parallel models by measuring the variance or the standard deviation of the outputs.
Figure~\ref{fig:bagging} schematically summarizes this idea.  If the physical results are reasonably constrained, we should anticipate that the outputs from these parallel models should be consistent with each other, which leads to small fluctuations and thus a small value of the standard deviation.  So far, this argument to identify the standard deviation in the bagging as the error bar is plausible; however, we must be careful of the interpretation as we discuss in what follows here.

In addition, it is more nontrivial how to take account of the error bar in the input side of observational data in a way that is desirable for the physics problem.  The natural tendency of the estimated error should be as follows: the more uncertain the input observational data is, the larger error bars the output theoretical parameters should have.  Below, we argue that it is not obvious whether this expected tendency of the error is satisfied or not.

\subsection{Subtleties in the previous method and a sketch of the new method}
\label{sec:outline.general}

\begin{figure}
    \centering
    \includegraphics[width=0.8\textwidth]{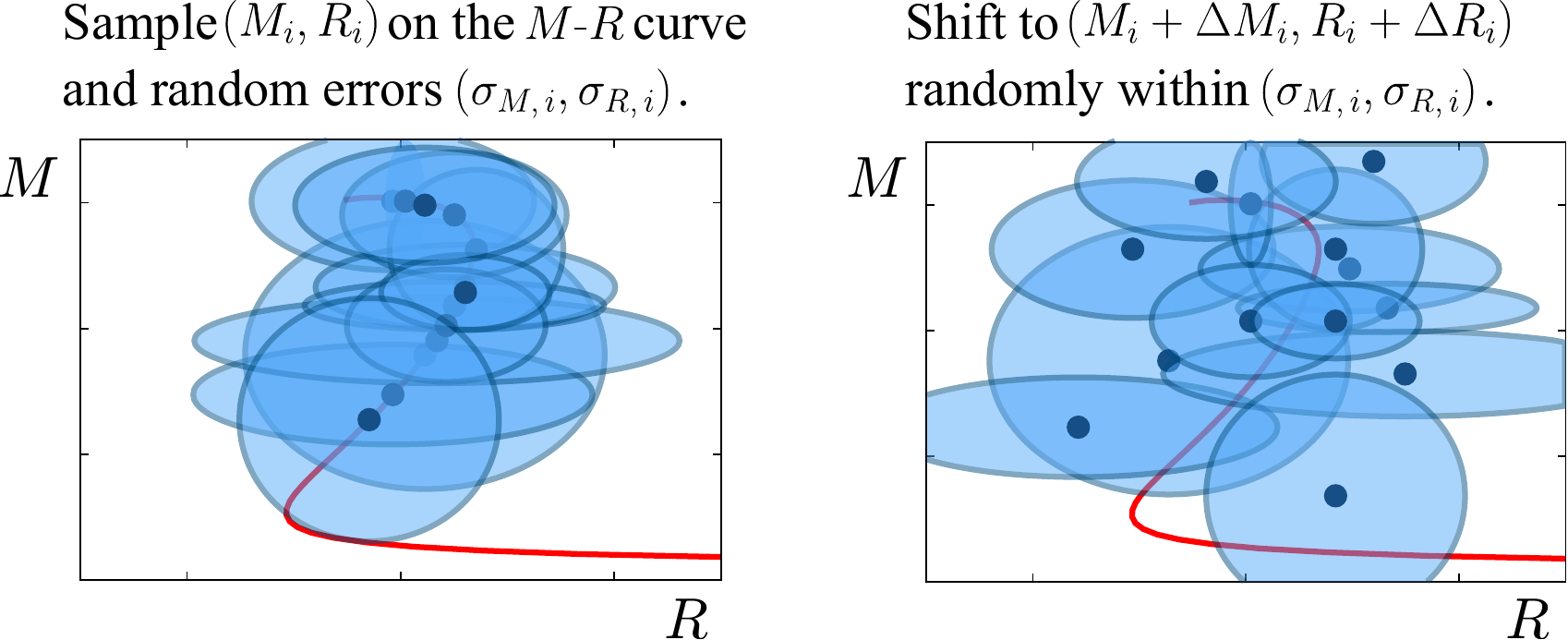}
    \caption{Generation of the training data in the previous works.  The set of $(M_i, R_i)$ is sampled on the $M$-$R$ curve obtained from the EoS with random errors $(\sigma_{M,i}, \sigma_{R,i})$ corresponding to the observation errors.  Then, randomly shifted data is generated to train the model so that the model can learn that the input data may have deviation within $(\sigma_{M, i}, \sigma_{R,i})$.}
    \label{fig:prev_noise}
\end{figure}

To clarify our motivation in this work, we shall explain the method we adopted in the previous works~\cite{Fujimoto:2017cdo,Fujimoto:2019hxv,Fujimoto:2021zas} more closely.
Figure~\ref{fig:prev_noise} shows how to take account of the observational uncertainty in preparing the training data for supervised learning.
First, we generate many EoSs randomly. We will explain the parametrization of the EoS in the next subsection.
Then, we solve the TOV equation to obtain a family of the NS mass $M$ and the radius $R$ as a function of the central pressure $\pc$, which is called the $M$-$R$ curve.
Next, we randomly sample $(M_i, R_i)$ on the $M$-$R$ curve as well as the errors $(\sigma_{M,\, i}, \sigma_{R,\, i})$ as sketched in the left panel of figure~\ref{fig:prev_noise}.
In reality, the data for each NS may have a far more non-trivial distribution than the Gaussian distribution, and we denote the real data distribution by $\PrObs$.
The ML model should ``know'' that the given data is subject to deviation from the genuine $M$-$R$ relation according to $\PrObs$.
Thus, as depicted in the right panel of figure~\ref{fig:prev_noise}, for each sampled point we augment the data by shifting $(M_i, R_i)$ to $(M_i+\Delta M_i, R_i+\Delta R_i)$ in accord with the Gaussian distribution with the standard deviations, $(\sigma_{M,\, i}, \sigma_{R,\, i})$.
In this way, we make noised copies of the data so that the ML model can be tolerant of uncertainty in the input and correctly guess the answer avoiding the over-fitting problem.
The error bar in the output is estimated by the bagging as schematically illustrated in figure~\ref{fig:bagging}.

Let us reiterate the above strategy in a symbolic way. Here, we assume that there are $\NSNumber (\in\mathbb{N})$ observed NSs, and we can generalize the mass and the radius as ``properties'' of each observation specified by $P (\in\mathbb{N})$ parameters such as the mass, the radius, the tidal deformability, the moment of inertia, etc.  Then, the observation data has $n_X:=\NSNumber P$ parameters in total and is collectively expressed as $\bm{X} = \{ (X_1^{(i)}, \dots, X_P^{(i)}) \}_{i=1}^{\NSNumber} \in \mathbb{R}^{n_X}$, where $X_p^{(i)}$ is the $p$-th property of the $i$-th NS\@.\footnote{Generally speaking, we can also take $\bm{X}$ as spectra from the actual telescope observation as discussed in refs.~\cite{Farrell:2022lfd, Farrell:2023ojk}.  In this case, however, the relation between the telescope spectra and the EoS is not as clear-cut as the relation between the $M$-$R$ curve and the EoS that is firmly based on general relativity.
Therefore, one needs to introduce the nuisance parameters in the former case and marginalization over these parameters.}
In the present work, specifically, $(X_1^{(i)}, X_2^{(i)})$ stands for the mass and the radius of the $i$-th NS, i.e., $(M^{(i)},\, R^{(i)})$.  As already discussed at the beginning of this section, astronomical observations do not directly give a point $\bm{X}$ because of observational uncertainties.  Thus, the input from the observation is inherently represented by a joint probability density $\PrObs(\bm{X})$ over $\mathbb{R}^{n_X}$.
In our previous works~\cite{Fujimoto:2019hxv, Fujimoto:2021zas}, we tried to characterize $\PrObs$ by the mean $\{(\mu_M^{(i)}, \mu_R^{(i)})\}_{i=1}^{\NSNumber}$ and the width $\{(\sigma_M^{(i)}, \sigma_R^{(i)})\}_{i=1}^{\NSNumber}$. 
We then proposed to build an NN model, $\Psi_\mathrm{prev}$, that directly predicts a likely EoS:
\begin{align}
  \begin{array}{cccc}
  \Psi_\mathrm{prev}:
  & \mathbb{R}^{n_X} \times \mathbb{R}^{n_X} & \to & \FuncSpace{\mathbb{R}} \\
  & \downIn && \downIn \\
  & (\bm{\mu}_{\bm{X}}, \bm{\sigma}_{\bm{X}}) & \quad \mapsto \quad & \bm{Y}
  \end{array}
  \label{eq:outline.previous-regression}
\end{align}
Here, the output EoS of our central interest is a relation, $p(\varepsilon)$, between the pressure $p$ and the energy density $\varepsilon$, which is collectively denoted by $\bm{Y} = p(\cdot) \in \FuncSpace{\mathbb{R}}$, where $\FuncSpace{\Omega}$ denotes the function space $\{\Omega\to \mathbb{R}\}$.

\begin{figure}
    \centering
    \includegraphics[width=0.7\textwidth]{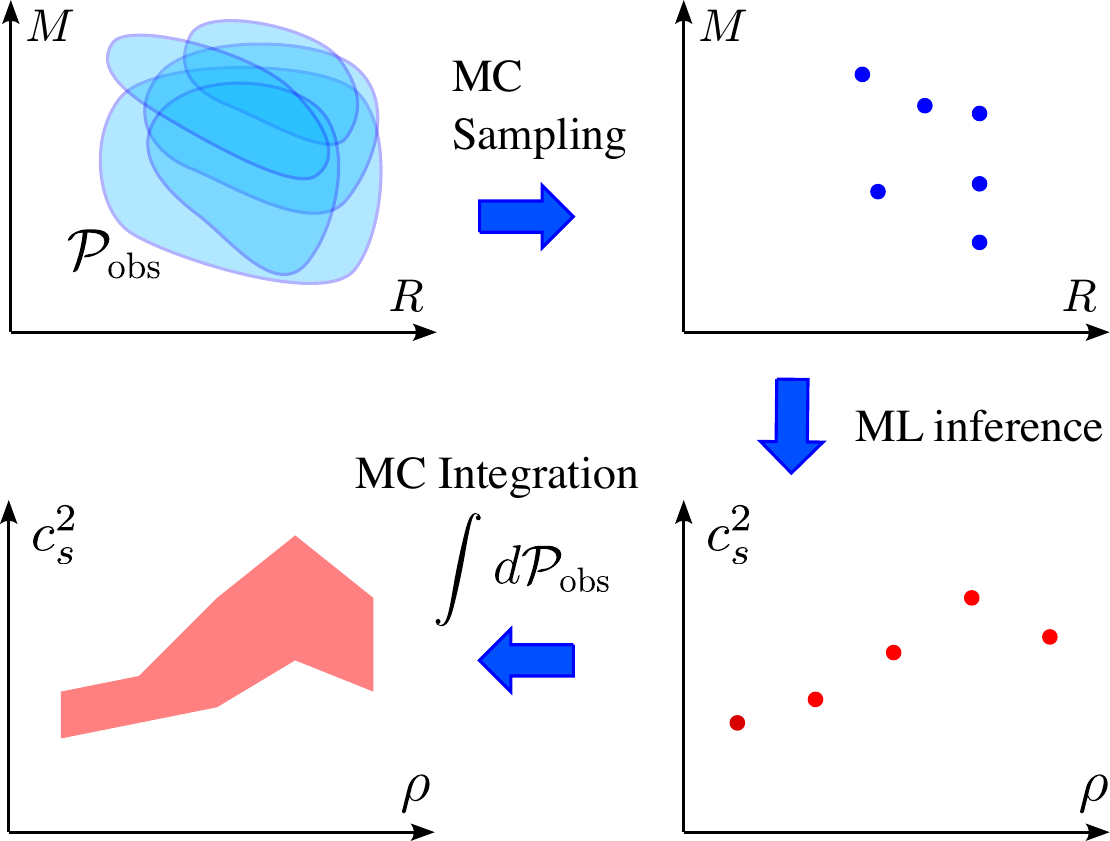}
    \caption{Our proposed procedures to deal with the error bars.  The observational data is supposed to have a distribution $\PrObs$ which spans on the $M$-$R$ plane in the present case (top-left).  Sets of data points are randomly sampled according to $\PrObs$ (top-right).  The ML model is trained to infer sets of the theoretical parameters, i.e., the sound velocities at various densities in the present case (bottom-right).  Finally, the ensemble average is taken with the weight $\PrObs$ on the input side, leading to the estimate for the probability distribution $\PrEoS$ on the output side (bottom-left).}
    \label{fig:Pobs}
\end{figure}

There are, however, several subtle points in this approach.
The observational data is not necessarily a simple product of the Gaussian distributions characterized by $(\sigma_{M,\,i},\sigma_{R,\,i})$ only.  The data for each NS could have a rather complicated distribution as shown by the shaded blobs in the top-left panel of figure~\ref{fig:Pobs}, which schematically represents $\PrObs$.
The problem is how to capture arbitrary shapes of the distribution $\PrObs$.  As sketched in figure~\ref{fig:Pobs} (and figure~\ref{fig:mr} for the real data), the distribution of the observational data may be significantly distorted and could not be well approximated by $(\sigma_{M,\,i},\sigma_{R,\,i})$.

There is another subtle problem, i.e., the uncertainty quantification in the final output.  To understand the problem better, let us assume a hypothetical parameter that is not constrained by the observational data at all.
If such a parameter exists, the uncertainty must be large.  In the actual estimate, however, such an unconstrained parameter could be dominantly chosen by the prior distribution, which is essentially the distribution in the training dataset in the present context.  For a fixed prior distribution, consequently, the standard deviation of the unconstrained parameter may happen to be small.
Roughly speaking, the undetermined output parameter may take close values that are far from the true answer.

The present paper focuses on the problem of the $\PrObs$ treatment.  In contrast, it is generally difficult to overcome the latter problem, and we should judiciously scrutinize the prior dependence.  In this paper, we will take a close look at different properties of the estimated uncertainties and defer the systematic investigation of the prior dependence to the future study.
Nonetheless, we will present exploratory analyses on the sensitivity of our results to selected prior choices.

Figure~\ref{fig:Pobs} is a cartoon to explain the new procedures we propose in the present work.  It is not our intention to claim that the present method supersedes the previous one.  Rather, we aim to establish an independent estimate for treating the error bars, which would be useful for the reliability check.  Here, for convenience, we use the terminology in a specific field of physics, but the problem of uncertainty quantification is ubiquitous, and it can be applied to a broad spectrum of problems.
In the top-left panel of figure~\ref{fig:Pobs}, $\PrObs$ represents the probability distribution of the observational data.  Given an ML model from the observational $M$-$R$ points to the theoretical EoS points, let us carry out the MC sampling of the $M$-$R$ data points following $\PrObs$.
For the points sampled on the $M$-$R$ plane as in the top-right panel of figure~\ref{fig:Pobs}, we can deduce the EoS parameters, which are depicted in the bottom-right panel of figure~\ref{fig:Pobs}.
Since the $M$-$R$ points are randomly sampled, we should take the ensemble average in the end, which is nothing but making the convolution with $\PrObs$.  We can finally reach the probability distribution of the EoS parameters denoted by $\PrEoS$, from which we can immediately read out the mean and the variance.

\subsection{Strategy of the present method}
\label{sec.outline.strategy}

We shall go into the details of the present method using the symbolic notation as used in the previous subsection.
Our goal is to make constraints on the NS EoS using the observations of static NS properties, for which we adopt only the mass $M$ and the radius $R$ in this work.  The problem is that the exact $\bm{Y}$ cannot be identified by the observational input because the observation data takes a distribution $\PrObs(\bm{X})$, and thus the corresponding EoS should naturally be given by a distribution as well. Even if the observational data comes along the genuine $M$-$R$ line without uncertainties, in principle, the observation of a \textit{finite} number of the NSs is generally insufficient to pin down the exact EoS in an \textit{infinite-dimensional} function space.  Therefore, we should consider the constraint to the NS EoS given in the form of a probability distribution, $\PrEoS(\bm{Y})$.
In the general term, we may express our ultimate goal as \textit{a regression $\mathscr{R}_{\mathrm{inv}}$ of ``a probability distribution of EoS'' $\PrEoS$ on ``a probability distribution of the NS properties'' $\PrObs$}:
\begin{align}
  \begin{array}{cccc}
  \mathscr{R}_{\mathrm{inv}}:
  & \FuncSpace{\mathbb{R}^{n_X}}
    & \to & \FuncSpace{\FuncSpace{\mathbb{R}}} \\
  & \downIn && \downIn \\
  & \PrObs(\cdot) & \quad \mapsto \quad & \PrEoS(\cdot)
  \end{array}
  \label{eq:outline.ultimate-regression}
\end{align}

One general and well-defined framework to perform this is the Bayesian analysis based on the MC integration, but it would be numerically demanding if we test the method in various observational scenarios and try to understand the characteristics of the method itself.  As an alternative approach, it would be interesting to think about a simpler ML-based setup.  

In this work, we examine a deep ML approach. Our central idea is to render the problem~\eqref{eq:outline.ultimate-regression} into the \textit{pointwise} regression with a supervised ML:
\begin{align}
  \begin{array}{cccc}
  \MapInv:
  & \bm{X} & \quad \mapsto \quad & \bm{Y} 
  \end{array}
  \label{eq:outline.point-regression}
\end{align}
In practice, because of the limitation of computational power, we shall restrict $\mathcal{F}(\mathbb{R})$ to an $N$-dimensional EoS subspace that spans over the parametrized form of $p(\varepsilon)$ with the EoS parameters $\bm{Y} \in \mathbb{R}^N$.
We represent $\MapInv$ by an NN model and train the model using a dataset of (input, output) pairs, $(\bm{X},\bm{Y})$, for supervised learning.
Once $\MapInv$ is obtained, we introduce the following Ansatz for the EoS distribution to model $\PrEoS(\bm{Y}) \simeq \PrEoS'(\bm{Y})$:
\begin{align}
  \mathscr{R}'_{\mathrm{inv}}: \quad
  \PrEoS'(\bm{Y}) &= \int \delta(\bm{Y} - \MapInv(\bm{X})) \PrObs(\bm{X}) \dif\bm{X}\,.
  \label{eq:outline.ansatz}
\end{align}
Using the distribution~\eqref{eq:outline.ansatz}, we will discuss the \textit{most likely} EoS and corresponding uncertainties.  Equation~\eqref{eq:outline.ansatz} is applicable for arbitrary shapes of the observational distribution $\PrObs$ without updating the trained model, $\MapInv$. Also, we can discuss the uncertainties of the resulting EoS using $\PrEoS'$.

\begin{figure}
    \centering
    \includegraphics[width=0.9\textwidth]{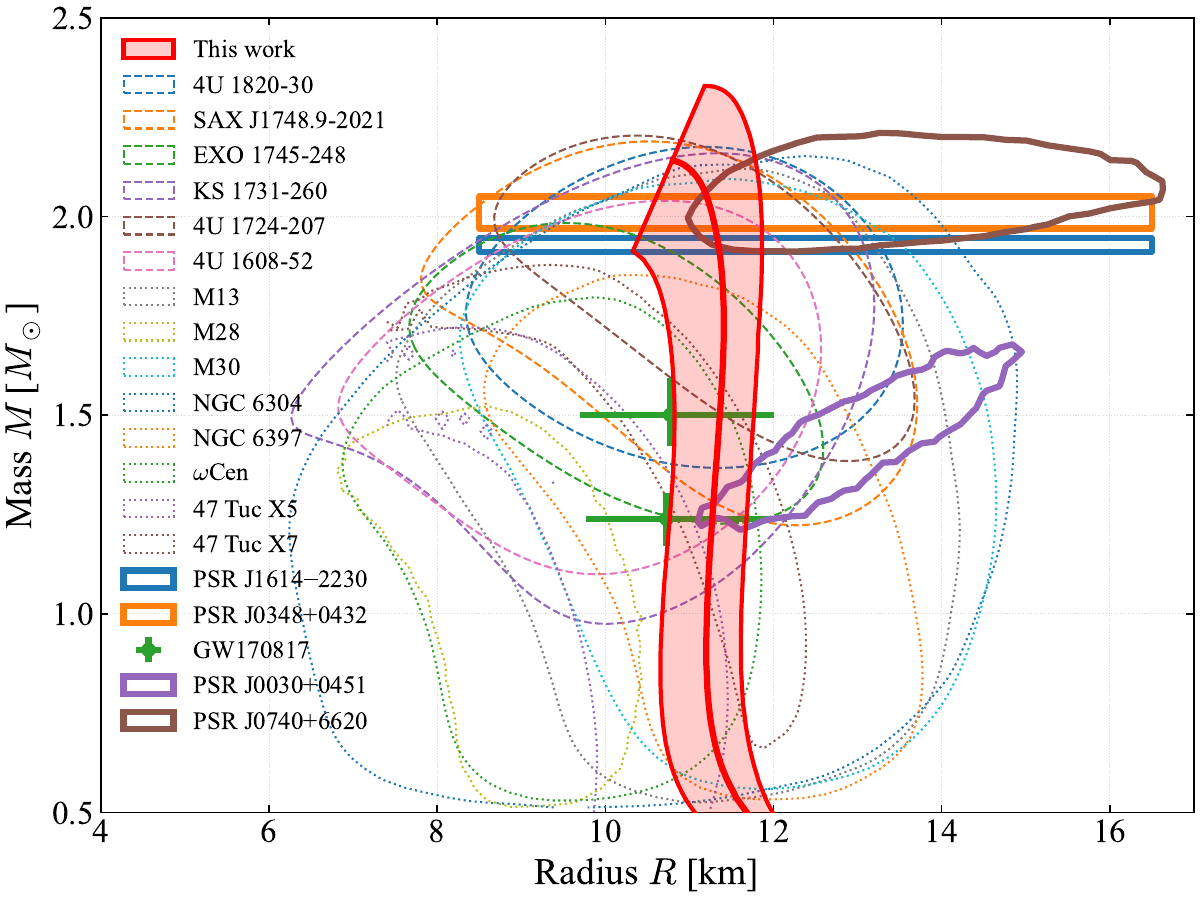}
    \caption{Distribution of the actual observational data.
    The $M$-$R$ region from the inferred EoS distribution (shown by the red band) is overlaid with the uncertainty estimate based on the method we propose in this work.}
    \label{fig:mr}
\end{figure}

For $\PrObs$, we use the collection of the data summarized in figure~\ref{fig:mr}.
These probability distributions are assembled from various sources~\cite{Ozel:2015fia, Bogdanov:2016nle, Demorest:2010bx, Antoniadis:2013pzd, Fonseca:2016tux, Miller:2019cac, Miller:2021qha, LIGOScientific:2018cki}.
In figure~\ref{fig:mr}, we plot the contour of the 68\% credible region of the posterior probability density function in Bayesian statistics except for the GW170817, J1614-2230, and J0348+0432 data.
The uncertainties of two NSs in the GW170817 data are intentionally shown by error bars because the distributions of the two NSs observed in a single binary NS merger event are highly correlated with each other through the tight constraint on the total mass of the system.
The rectangular bands for the J1614-2230 and J0348+0432 data show that the $R$-direction is uniformly distributed while the $M$-direction is limited within the 68\% confidence interval.
The NSs marked with thick lines in figure~\ref{fig:mr} are the data not incorporated in the previous analysis~\cite{Fujimoto:2019hxv} and newly considered in this work.  The red band is the $M$-$R$ curve with uncertainties corresponding to the EoS predicted by our NN method as shown in figure~\ref{fig:eos_result}), which constitutes the main results of this paper.
The details will be elaborated in section~\ref{sec:results}.

\section{Mapping $M$-$R$ distributions to EoS using neural network}
\label{sec:map}

We summarize the setup of our inference to predict the EoS parameters, $\bm{Y}$, from the NS properties, $\bm{X}$.

\begin{figure}
  \centering
  \includegraphics[width=0.9\textwidth]{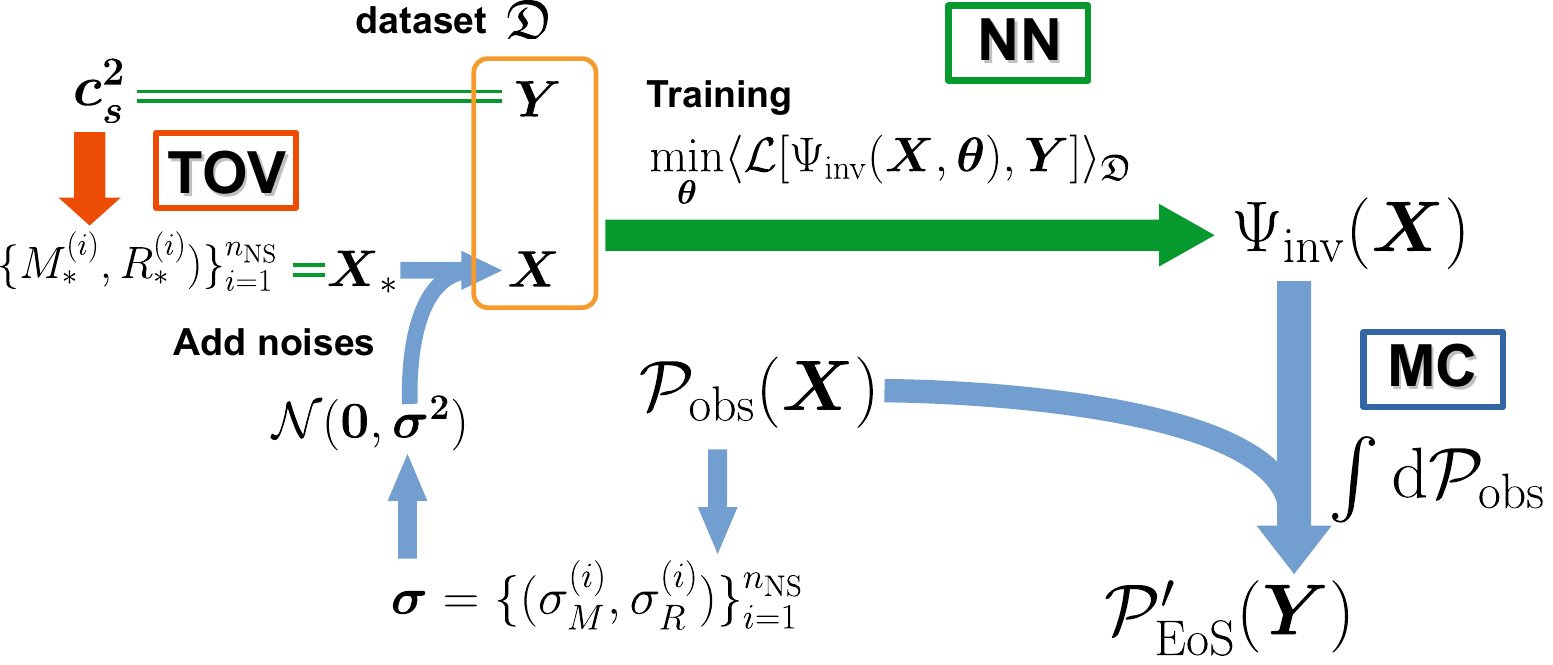}
  \caption{Schematic figure of our strategy to infer the distribution, $\PrEoS\simeq \PrEoS'$ of the EoS parameters, $\bm{Y} = {\bm c_s^2}$ (and $p_c$), from the experimental distribution, $\PrObs(\bm{X})$.}
  \label{fig:strategy}
\end{figure}

\subsection{Synopsis of our method}

Our strategy is illustrated in figure~\ref{fig:strategy}, and we describe our method along this schematic figure in this subsection.

We consider $\bm{X} = \{ (M^{(i)}, R^{(i)}) \}_{i=1}^{\NSNumber} \in \mathbb{R}^{2\NSNumber}$ for the NS properties, where $M^{(i)}$ and $R^{(i)}$ are the mass and the radius, respectively, of the $i$th NS\@.
For the EoS parameters $\bm{Y}$, we consider a piecewise polytropic form of the EoS and represent the EoS with a set of the speed of sound, $\bm{Y} := \bm{c_s^2} = \{c_{s,n}^2 \}_{n=1}^N$ (see section~\ref{sec:EoS} for detail).
The joint probability distribution $\PrObs(\bm{X})$ is specified by the observational data itemized in section~\ref{sec:obs}.

We first generate the dataset, $\mathfrak{D} = \{(\bm{X}_{i}, \bm{Y}_{i})\}_{i=1}^I$, for supervised learning (see the yellow box on the top left corner in figure~\ref{fig:strategy}).
We randomly sample EoS parameters $\bm{c_s^2} = \bm{Y}$.
The corresponding NS properties $\bm{X}$ are obtained by adding noises to the solutions of the TOV equation $\bm{X}_\ast$ (the red arrow on the top left corner).
The noises are sampled from the normal distribution $\mathcal{N} (0, (\sigma_M^{(i)})^2)\times \mathcal{N}(0, (\sigma_R^{(i)})^2)$, and the standard deviations $\{(\sigma_M^{(i)}, \sigma_R^{(i)})\}$ are determined so as to match those of $\PrObs(\bm{X})$ (the bottom left corner of figure~\ref{fig:strategy}).
Further details on the generation of the dataset $\mathfrak{D}$ will be elaborated in section~\ref{sec:method.dataset}.
We then design the NN architecture of $\MapInv$ (explained in section~\ref{sec:Arch}) and train it with the dataset $\mathfrak{D}$ (the green arrow in figure~\ref{fig:strategy}). 
We finally analyze the ``posterior'' distribution $\PrEoS'$ obtained by the MC integration over the $\bm{X}$-space convoluted with $\PrObs(\bm{X})$ as given in eq.~\eqref{eq:outline.ansatz} (the bottom right corner of figure~\ref{fig:strategy}).

To infer a ``likely'' EoS using $\PrEoS'$, we consider the Bayes estimator of the EoS parameters: $\bm{Y}^* := \langle \bm{Y}\rangle_{\PrEoS'}$, where $\langle\cdot\rangle_{\PrEoS'}$ denotes the average by the distribution $\PrEoS'$.
We then estimate its uncertainties by the covariances $\langle (\bm{Y} - \bm{Y}^*)(\bm{Y} - \bm{Y}^*)^\mathrm{T}\rangle_{\PrEoS'}$.
Instead of explicitly constructing $\PrEoS'$, we perform the MC integration to directly evaluate the averages $\langle\cdot\rangle_\mathrm{\PrEoS'}$:
\begin{align}
  \langle \mathcal{O}[\bm{Y}]\rangle_{\PrEoS'}
  &= \int \mathcal{O}[\MapInv(\bm{X})] \PrObs(\bm{X}) \dif\bm{X}
  \simeq \frac1{n_\mathrm{MC}} \sum_{i = 1}^{n_\mathrm{MC}} \mathcal{O}[\MapInv(\bm{X}_{i})]\,,
  \label{eq:uncertainty-forwarding}
\end{align}
where $\mathcal{O}[\bm{Y}]$ is a quantity written in terms of the EoS parameters $\bm{Y}$.
Also, $\bm{X}_{i}$ is the $i$th MC sampling point following the distribution $\PrObs(\bm{X})$,
and $n_\mathrm{MC}$ is the number of the sample points.

\subsection{EoS parametrization}
\label{sec:EoS}

In principle, an EoS can be defined through the speed of sound $c_s^2(\varepsilon) = \partial p/\partial \varepsilon$ as $p(\varepsilon) = \int_0^\varepsilon c_s^2(\varepsilon') d\varepsilon'$.
In this study, we employ the piecewise polytropic expression for an EoS, and it is parametrized by a set the speed of sound, $\bm{c_s^2} = \{c_{s,n}^2 \}_{n=1}^N$.
We split the range of considered density $[\varepsilon_0, \varepsilon_N]$ into $N$ ($\in \mathbb{N}$) segments, i.e., $[\varepsilon_{n-1}, \varepsilon_n]$ ($n = 1, \dots, N$), and specify a value of the average speed of sound: $c_{s,n}^2 = \int_{\varepsilon_{n-1}}^{\varepsilon_n} c_s^2(\varepsilon') d\varepsilon'/(\varepsilon_n - \varepsilon_{n-1})$ in each segment.
The density range is equally partitioned into segments in the logarithmic scale as
\begin{align}
  \ln\varepsilon_n := \ln\varepsilon_0 + \frac nN (\ln \varepsilon_N - \ln \varepsilon_0) \qquad (n = 1, \dots, N)\,,
\end{align}
where we chose $\varepsilon_0 = 150~\text{MeV}/\text{fm}^3$ to approximate the energy density at the nuclear saturation density $n_B = n_0$, and we set $\varepsilon_N := 8\varepsilon_0$.
For instance, a choice of $N=5$ leads to $(\varepsilon_0, \varepsilon_1, \varepsilon_2, \varepsilon_3, \varepsilon_4, \varepsilon_5) \simeq (1.0, 1.5, 2.3, 3.5, 5.2, 8.0) \varepsilon_0$.
The low-density EoS, $p(\varepsilon<\varepsilon_0)$, is constrained by existing models of nuclear matter, for which we employ SLy4~\cite{Douchin:2001sv}.
The pressure $p_n$ at the segment boundary $\varepsilon_n$ is obtained as
\begin{align}
  p_n
  &:= p_{n-1} + \int_{\varepsilon_{n-1}}^{\varepsilon_n} c_s^2(\varepsilon') d\varepsilon' \notag \\
  &= p_{n-1} + c^2_{s,n} (\varepsilon_n - \varepsilon_{n-1}) \qquad  (n = 1,\dots, N)\,,
  \label{eq:eos.piecewise-pressure}
\end{align}
where $p_0=p(\varepsilon_0)$ is fixed by SLy4. 
The EoS inside each segment is interpolated by a polytrope function, i.e., $p=p_{\mathrm{pol}}(\varepsilon) = K_n \varepsilon^{\Gamma_n}$ for $\varepsilon_{n-1} < \varepsilon < \varepsilon_n$,
where $(\Gamma_n,K_n)$ is determined by the boundary conditions, $p_{n-1} = p_{\mathrm{pol}}(\varepsilon_{n-1})$ and $p_{n} = p_{\mathrm{pol}}(\varepsilon_{n})$, which yields:
\begin{align}
  \Gamma_n = \frac{\ln p_{n} - \ln p_{n-1}}{\ln \varepsilon_{n} - \ln \varepsilon_{n-1}}\,, \qquad
  K_n
  &= p_n^{1/(1 - \ln\varepsilon_n/\ln\varepsilon_{n-1})} p_{n-1}^{1/(1 - \ln\varepsilon_{n-1}/\ln\varepsilon_n)}\,.
  \label{eq:eos.piecewise-parameter}
\end{align}
In this paper, we present results for $N=5$ unless otherwise specified.

\subsection{Dataset generation}
\label{sec:method.dataset}
\label{sec:TOV}

Here, we describe the procedure to generate the dataset used for training and validation.
Specifically, we take the following steps to prepare our dataset,
which is mostly in line with ref.~\cite{Fujimoto:2021zas} except for step 4 in the following:
\begin{enumerate}
\item We first generate an EoS, $\bm{Y} = \{c_{s,n}^2\}_{n=1}^N$,
  by randomly sampling each $c_{s,n}^2$ by the uniform distribution whose support is $[\delta, 1 - \delta]$ with $\delta = 0.01$.
\item We then obtain the corresponding $M$-$R$ line by solving the TOV equations described below.
  If the maximum mass $M_\mathrm{max}$ does not reach $2.01 M_{\odot}$, we reject the EoS\@.
  If the EoS is accepted, we proceed to the next step.
\item We randomly sample $\NSNumber$ NSs, $\{(M_*^{(i)}, R_*^{(i)})\}_{i=1}^{\NSNumber}$,
  on the $M$-$R$ line by the uniform distribution with respect to $M_*^{(i)}$ in the range $[M_\mathrm{min}, M_\mathrm{max}]$,
  where we take $M_\mathrm{min} = 1.0 M_{\odot}$.
\item We sample $\NSNumber$ observational data points, $\bm{X} = \{(M^{(i)}, R^{(i)})\}_{i=1}^{\NSNumber}$, from the Gaussian distributions:
  $(M^{(i)}, R^{(i)}) \sim \mathcal{N}(M_*^{(i)}, (\sigma_{M}^{(i)})^2) \times \mathcal{N}(R_*^{(i)}, (\sigma_{R}^{(i)})^2)$.
  Here, the standard deviations, $\{(\sigma_{M}^{(i)}, \sigma_{R}^{(i)})\}_{i=1}^{\NSNumber}$, are matched with those of $\PrObs(\bm{X})$.
\item A pair of $(\bm{X}, \bm{Y})$ forms one data point.  We repeat steps 1--4 to generate many pairs to create a dataset.
    To increase the dataset size, we may sample multiple sets of $\NSNumber$ NSs for a single step in step 3,
    and multiple sets of observational points for a single set of $\NSNumber$ NSs in step 4.
\end{enumerate}

We note that in the previous method~\cite{Fujimoto:2019hxv, Fujimoto:2021zas}, at step~4, we had to sample the standard deviations, $\{(\sigma_{M}^{(i)}, \sigma_{R}^{(i)})\}_{i=1}^{\NSNumber}$, randomly by the uniform distributions denoted by $\UniDis(0,M_\odot)$ and $\UniDis(0,5\,\mathrm{km})$.  For step 2,
$M$ and $R$ of NSs can be obtained from the EoS
through the TOV equation~\cite{Tolman:1939jz, Oppenheimer:1939ne}.
The TOV equation describes the hydrostatic equilibrium condition of spherically symmetric configuration in general relativity, which takes the following form:
\begin{align}
  \frac{\dif{p}(r)}{\dif{r}} &= - \frac{m(r) \varepsilon(p(r))}{r^2} \left[1 + \frac{p(r)}{\varepsilon(p(r))}\right]
  \left[1 + \frac{4 \pi r^3 p(r)}{m(r)}\right]\left[1-\frac{2m(r)}{r}\right]^{-1}\,, \label{eq:TOVeq} \\  \frac{\dif{m}(r)}{\dif{r}} &= 4 \pi r^2 \varepsilon(p(r))\,, \label{eq:TOVeq2}
\end{align}
where $r$ is the radial distance from the NS center,
$p(r)$ is the pressure inside the NS at $r$,
and $m(r)$ is the mass enclosed within the sphere of radius $r$.
We can numerically solve eqs.~\eqref{eq:TOVeq} and~\eqref{eq:TOVeq2}
with the boundary conditions,
\begin{align}
\label{eq:pc}
  p(r=0) = p_c\,, \qquad m(r=0) = 0,
\end{align}
using a standard method for ordinary differential equations outward from $r=0$ until the pressure vanishes.
The NS radius, $R$, is defined by the condition of vanishing pressure, and the NS mass, $M$, is defined accordingly as
\begin{align}
  R:\ p(R) = 0\,, \qquad M := m(R)\,.
\end{align}
Therefore, the resulting pair of $M$ and $R$ depends on the parameter $p_c$ in eq.~\eqref{eq:pc}.  We can draw the $M$-$R$ curve by changing the parameter $p_c$.

\subsection{Neural-network models}
\label{sec:Arch}

To represent the fitting function $\MapInv(\bm{X})$, we employ a fully connected deep neural network.
The input $\bm{X} \in \mathbb{R}^{d_\mathrm{in}}$ is normalized for pre-processing:
\begin{align}
  \bar{M}^{(i)} := \frac{M^{(i)}}{M_{\rm norm}}\,,\qquad 
  \bar{R}^{(i)} := \frac{R^{(i)}}{R_{\rm norm}} \qquad
  (i = 1, \dots, \NSNumber)\,,
\end{align}
with $M_\mathrm{norm} = 3 M_{\odot}$ and $R_\mathrm{norm} = 30~\text{km}$ in our choice.
The normalized NS properties are plugged into the input layer: $\bm{x}^{(0)}=\bm{X}$.
The NN model iteratively calculates the values of the subsequent layers as
\begin{align}
  x_i^{(l)} = a^{(l)}\biggl(\sum_{j=1}^{n_{l-1}} W^{(l)}_{ij} x_j^{(l-1)} + b_i^{(l)}\biggr) \qquad (l = 1, \dots, n_{\mathrm{layer}})\,,
\end{align}
where $\bm{x}^{(l)} \in \mathbb{R}^{d^{(l)}}$ is the state of the $l$-th layer with $d^{(l)}$ being the number of neurons in the $l$-th layer.
The activation function for the $l$-th layer is $a^{(l)}(x)$,
and $W_{ij}^{(l)}$ and $b_j^{(l)}$ are the trainable parameters.
The output in the last layer gives $\bm{Y} = \bm{x}^{(n_\mathrm{layer})}$, which is the EoS parameters, $\bm{Y} = \bm{c_s^2}$, in our model.
It should be noted that our extended analysis in section~\ref{sec:pc} will instead adopt $\bm{Y} = (\bm{c_s^2},\{p_c^{(i)}\}_{i=1}^{\NSNumber})$ to additionally predict the central pressure $p_c^{(i)}$ of the $i$-th NS\@.
For the motivation of analysis with and without $p_c$, we will closely discuss it later in section~\ref{sec:pc}.
The NN structure used in this work is summarized in table~\ref{table:NNarc}.
The dimensions of the input and output and the number of trainable parameters of the NN models in our analysis are summarized in table~\ref{table:num_of_params}, which are estimated from $\dim[W^{(l)}] = d^{(l-1)} d^{(l)}$ and $\dim[b^{(l)}] = d^{(l)}$.

\begin{table}
 \centering
  \begin{tabular}{|c|cccccc|}
   \hline
   Layer & $l=0$ (Input) & $l=1$ & $l=2$ & $l=3$ & $l=4$ & $l=n_{\mathrm{layer}}$ (Output) \\
   \hline
   $d^{(l)}$ &  $d_{\rm in}$ & $4 d_{\rm in}$ & $4 d_{\rm in}$ & $2 d_{\rm in}$ & $2 d_{\rm in}$ & $d_{\rm out}$ \\
   $a^{(l)}$ & $-$ & ReLU & ReLU & ReLU &  ReLU & Sigmoid \\
   \hline
  \end{tabular}
  \caption{Our NN architecture.
  The constants $d_{\rm in/out}$ represent the dimensions (the number of neurons) of the input/output layer.
  }
  \label{table:NNarc}
\end{table}

\begin{table}
 \centering
  \begin{tabular}{|lc|ccc|}
  \hline
  & & $\NSNumber=14$ & $\NSNumber=6$ & $\NSNumber=20$ \\
  \hline\hline
   Model w/o $p_c$ & $(d_{\rm in},\, d_{\rm out})$ & (28,\,5) & (12,\,5) & (40,\,5) \\
   & \# params & 25,709 & 4,877 & 52,085 \\
   \hline
   Model with $p_c$ & $(d_{\rm in},\, d_{\rm out})$ & (28,\,19) & (12,\,11) & (40,\,25) \\
   & \# params & 26,507 & 5,027 & 53,705 \\
   \hline
  \end{tabular}
  \caption{
    The numbers of trainable parameters. 
    In the input layer, $d_{\rm in}=2\NSNumber$ corresponds to our choice of $M$ and $R$ for the NS properties.  In the output layer, $d_{\rm out}=N$ for the model without $p_c$ and $d=N+\NSNumber$ for the model with $p_c$.
    We take $N=5$ and $\NSNumber=14$, $6$, and $20$ (see the text for details).
    }
  \label{table:num_of_params}
\end{table}

For our numerics, we use the \texttt{Keras} package~\cite{software:Keras} for the NN construction and employ $\texttt{Adam}$~\cite{DBLP:journals/corr/KingmaB14} and $\texttt{msle}$ for the optimizer and the loss function, respectively.
We generate $200,000$ independent $M$-$R$ lines, repeating 20 times of sampling $\NSNumber$ observational data points (corresponding to step~3 in section~\ref{sec:method.dataset}) per one generated $M$-$R$ line, and then repeat 10 times of augmenting the data points with Gaussian noises per one set of the data points (corresponding to step~4 in section~\ref{sec:method.dataset}).
Hence, we prepare the dataset consisting of $200,000 \times 20 \times 10 = 40,000,000$ observation data.  We split it into $40,000,000 \times 0.9 = 36,000,000$ and $40,000,000 \times 0.1 = 4,000,000$ observation data as the datasets used for training and validation, respectively.
We take $200$ epochs with the batch size of $100,000$ for the optimization\footnote{
  We checked that reducing the batch size to 10,000, for example, does not change the results within the error bars.
}.

\subsection{Observed properties of neutron stars}
\label{sec:obs}

We explain how to prepare $\PrObs(\bm{X})$ in the present analysis in what follows in this subsection.
All the sources of data are given in the form of the posterior probability density function in Bayesian statistics;
see figure~\ref{fig:mr} for the 68\% contour of their credible regions.

We use the following 14 sources of spectral measurement in X-ray, which will be collectively referred to as ``PREs/qLMXBs'' hereafter.
They were also used in our previous analyses~\cite{Fujimoto:2019hxv, Fujimoto:2021zas}.
\begin{itemize}
\item Photospheric radius expansion X-ray bursts (PREs):
4U 1608-52, 4U 1724-207, KS 1731-260, EXO 1745-248, SAX J1748.9-2021, and 4U 1820-30~\cite{Ozel:2015fia} ($\mbox{\# distributions} = 6$)
\item Quiescent low-mass X-ray binaries (qLMXBs):
M13, M28, M30, NGC 6304, NGC 6397, $\omega$ Cen~\cite{Ozel:2015fia}, X5, and X7~\cite{Bogdanov:2016nle}
($\mbox{\# distributions} = 8$)
\end{itemize}

On top of them, we further incorporate the following 5 new data into the analyses in this work\footnote{Precisely speaking, the NICER data from J0030+0451 was already taken into account in our previous analysis~\cite{Fujimoto:2021zas}.}, which will be collectively referred to as ``HP/NICER/GW'' later.
\begin{itemize}
\item Heavy pulsars (HP) from precise radio astronomy:
  PSR J1614-2230~\cite{Demorest:2010bx, Fonseca:2016tux}, and PSR J0348+0451~\cite{Antoniadis:2013pzd} ($\mbox{\# distributions} = 2$)
\item Radii from the timing measurements in the X-ray band in NICER:
  PSR J0030+0451~\cite{Miller:2019cac}, and PSR J0740+6620~\cite{Miller:2021qha} ($\mbox{\# distributions} = 2$)
\item Gravitational wave event from LIGO-Virgo collaboration (GW):
  GW170817~\cite{LIGOScientific:2018cki}\\ ($\mbox{\# distributions} = 1$)
\end{itemize}
The heavy pulsars are the precise radio measurements of masses of the heaviest NS known to date except for the black widow pulsar~\cite{Romani:2022jhd}.
We did not adopt the black widow pulsar data because of a larger uncertainty compared to the two pulsars listed above.
We limited the radii of the heavy pulsars within $8.5$--$16.5~\text{km}$.
The edge values correspond to the smallest and largest possible radii~\cite{Drischler:2020fvz}, respectively, extrapolated from the crust EoS assumed up to half the saturation density.

We define $\PrObs(\bm{X})$ by a product of all those distributions.
It should be noted that two pairs of $(M,R)$ from GW170817 in the binary system of two NSs are correlated to each other.
Hence, the numbers of NSs for the two categories are $n_{\mbox{\tiny PREs/qLMXBs}} = 14$ and $n_{\mbox{\tiny HP/NICER/GW}} = 6$, respectively, and the total number amounts to $\NSNumber=20$.

\section{Results for the Equation of State} \label{sec:results}
Our main results of the EoS, in particular the behavior of $c_s^2$, are discussed in this section.

\begin{figure}
  \centering
  \includegraphics[width=0.9\textwidth]{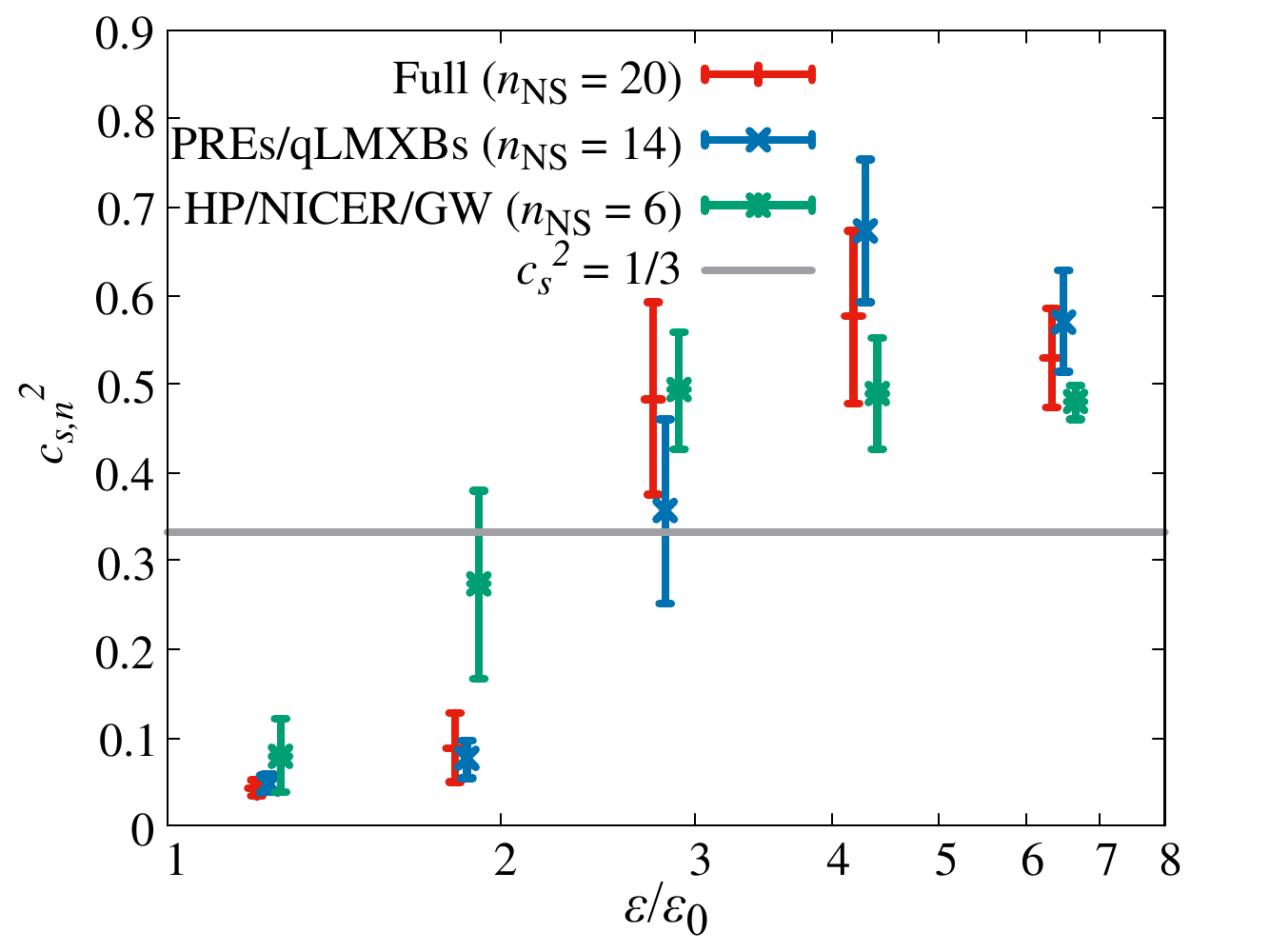}
  \caption{Speed of sound squared, $c_{s,n}^2$, as a function of the energy density. The results inferred from the full, PREs/qLMXBs, and HP/NICER/GW distributions are shown by the red, blue, and green bars, respectively, which represent the standard deviation of $\bm{Y}$ over $\PrEoS'$.}
\label{fig:MR_uniM_cs2only}
\end{figure}

Figure~\ref{fig:MR_uniM_cs2only} shows $c_{s,n}^2$ estimated by the full distribution of $\PrObs$. We also compare it with the results from partial distributions limited to only the PREs/qLMXBs and the HP/NICER/GW data.
The mean values $\mu[\bm{Y}]$ and the standard deviation $\sigma[\bm{Y}]$ are obtained by
\begin{align}
\mu[Y_i] & = \langle Y_i \rangle_{\PrEoS'}\,, \qquad  \sigma[{Y}_i]  = \sqrt{\langle (Y_i  - \mu[Y_i])^2\rangle_{\PrEoS'}}\,.
\label{eq:mean_error} 
\end{align}
The behavior of PREs/qLMXBs significantly differs from that of HP/NICER/GW\@.
As seen in figure~\ref{fig:MR_uniM_cs2only}, the PREs/qLMXBs data generates a soft EoS with $c_s^2 \sim 0.1$ in the low $\varepsilon$ regime and the EoS starts to be stiff only around the middle segments at $\varepsilon/\varepsilon_0 \gtrsim 2$, whereas the EoS inferred from the HP/NICER/GW data starts to be stiff already in the second segment;  the latter EoS has a larger sound speed around $\varepsilon / \varepsilon_0 \simeq 2$ compared to the typical value, e.g., the ab initio calculation from the chiral effective theory~\cite{Drischler:2020fvz, Drischler:2020hwi} although they are still consistent within the uncertainty.

The high-density part of the EoS inferred from the NP/NICER/GW data, particularly the fifth segment, may be unconstrained by the data as it is apparently set by the average of the prior distribution $c_{s,5}^2 \simeq 0.5$.
Further, the small uncertainty signifies that this may be the biased output from the NN and does not necessarily reflect the actual statistical errors.
We will come back to this point later in section~\ref{sec:pc} when we include the central pressure of NSs in the analysis.

\begin{figure}
  \centering
  \includegraphics[width=0.9\textwidth]{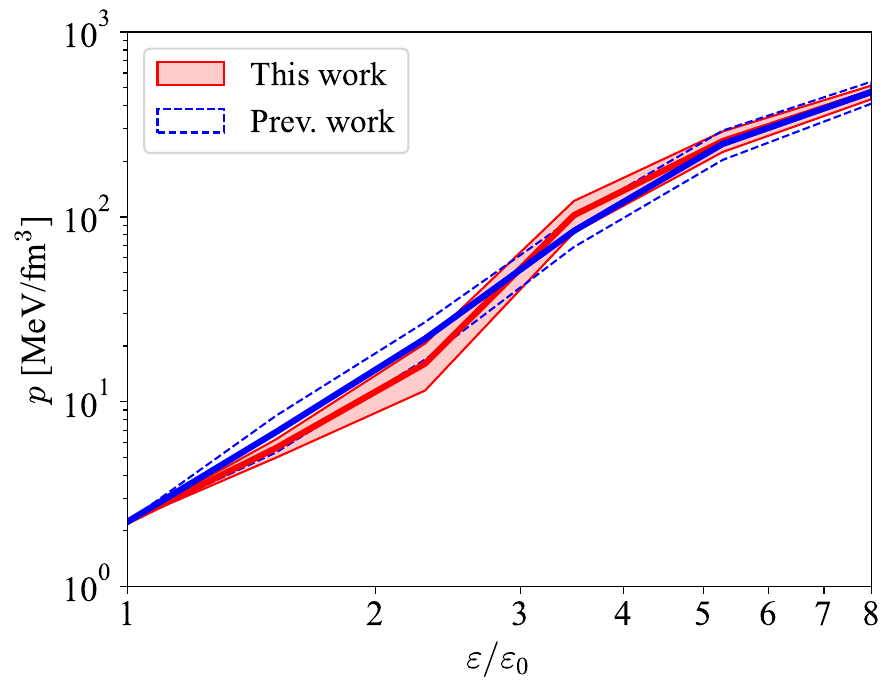}
  \caption{EoS with the uncertainty estimate based on the method we propose in this work. The colored lines and bands are the mean values and the errors, respectively. The upper and lower lines of the bands are obtained by substituting $\bm{c_s^2} = \mu[\bm{Y}] \pm \sigma[\bm{Y}]$ in eqs.~\eqref{eq:eos.piecewise-pressure} and~\eqref{eq:eos.piecewise-parameter}.}
\label{fig:eos_result}
\end{figure}

In figure~\ref{fig:eos_result}, we show the pressure $p$ obtained from the constrained $c_{s,n}^2$. The pressure is calculated as a piecewise polytropic form of a function of $\varepsilon$ by substituting $\bm{c_s^2} = \mu[\bm{Y}]$ in eqs.~\eqref{eq:eos.piecewise-pressure} and~\eqref{eq:eos.piecewise-parameter}.
The EoS marked as ``This work'' in figure~\ref{fig:eos_result} corresponds to $c_{s,n}^2$ marked as ``Full'' in figure~\ref{fig:MR_uniM_cs2only}.
We compared our EoS results with our previous estimates from ref.~\cite{Fujimoto:2019hxv}.
The inferred EoS in this work exhibits a slightly steeper transition from a soft EoS at low density to a stiff EoS at middle density compared to that in the previous work.

\begin{figure}
    \centering
    \begin{tabular}{cc}
      \begin{minipage}{0.5\textwidth}
        \centering
          \includegraphics[clip, width=0.85\textwidth]{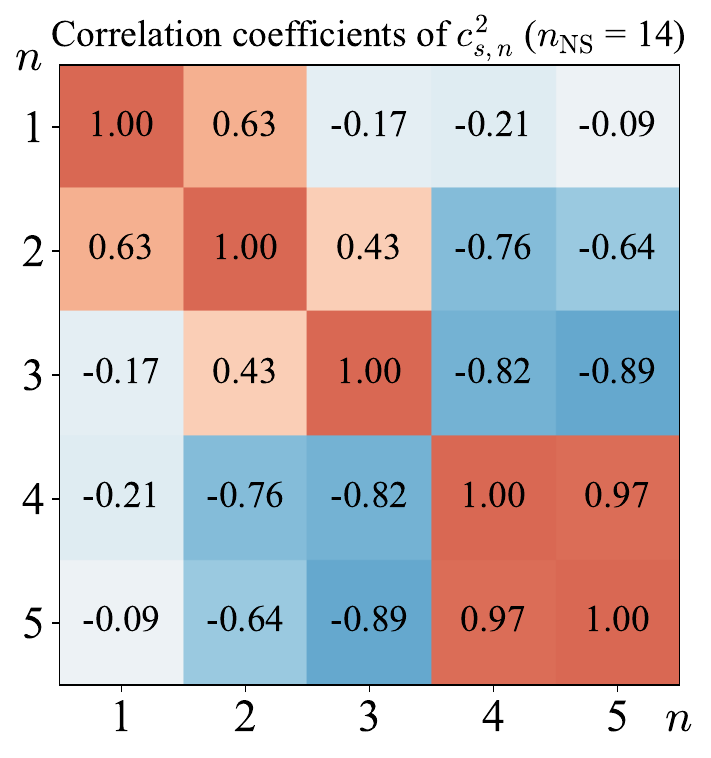}\\
          (a) PREs/qLMXBs
      \end{minipage}
      \begin{minipage}{0.5\textwidth}
        \centering
          \includegraphics[clip, width=0.85\textwidth]{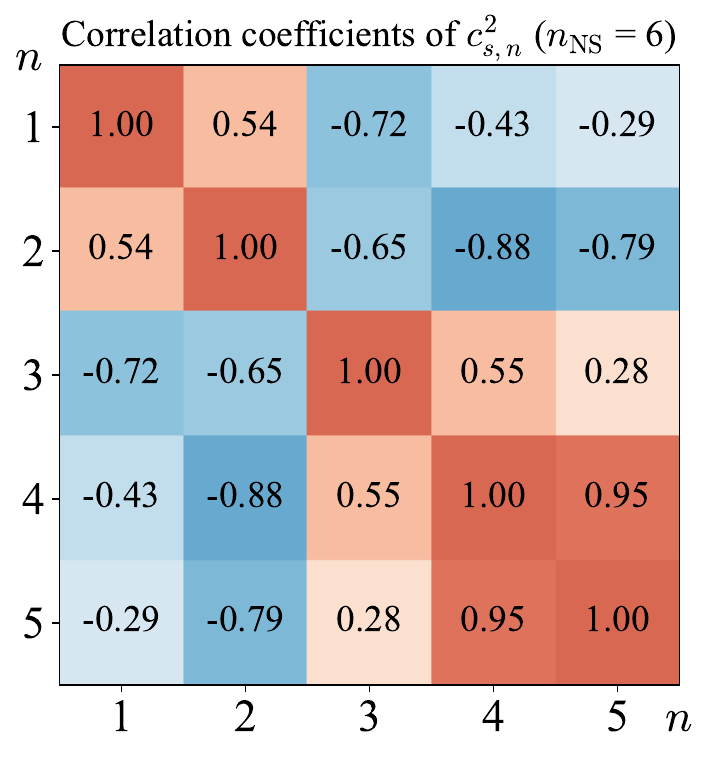}\\
          (b) HP/NICER/GW
      \end{minipage}
      \vspace{1em}\\
      \begin{minipage}{0.5\textwidth}
        \centering
          \includegraphics[clip, width=0.85\textwidth]{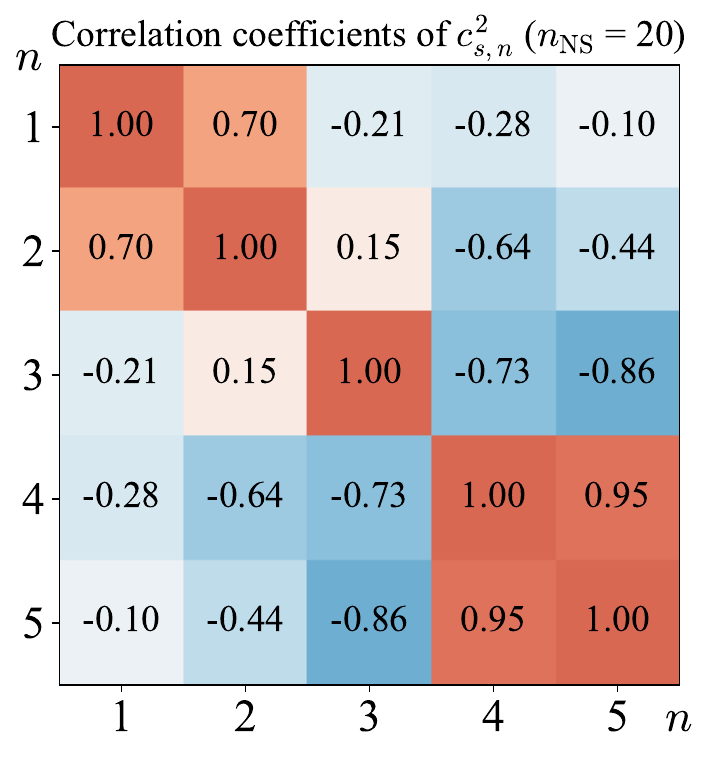}\\
          (c) Full 
      \end{minipage}
    \end{tabular} 
    \caption{Correlation coefficients of $c_{s,n}^2$ ($n=1,\dots,5$) for $\NSNumber=14$ (PREs/qLMXBs), $6$ (HP/NICER/GW), and $20$ (Full distributions).
    The red and blue colored cells correspond to positive and negative values, respectively.
    The diagonal components are unity by definition of eq.~\eqref{eq:covariance_mat}.
    }
    \label{fig:covmat}
\end{figure}

The EoS presented by the thick solid line in figure~\ref{fig:eos_result} is based on the point estimation,
but it is just one possibility among various EoSs.
As already discussed in sections~\ref{sec:introduction} and~\ref{sec:outline}, it is crucial to estimate the uncertainty in constraining physics corresponding to the uncertainty in observational data.  The standard deviation over $\PrEoS'$ is presented by the band in figure~\ref{fig:eos_result} as one measure of the uncertainty.
Moreover, it should be noted that those estimated values of $c_{s,n}^2$ at different $n$ are not independent of each other.
They can be correlated with each other through the nontrivial shape of $\PrEoS'$ generated by $\PrObs$.
Such correlation affects the shape of the possible EoSs.
As the simplest measure for the correlation among $c_{s,n}^{2}$'s,
we consider the Pearson correlation coefficients:
\begin{align}
r_{ij} & := \frac{\langle (Y_i - \mu[Y_i]  ) (Y_j -  \mu[ Y_j] )\rangle_{\PrEoS'}}{\sigma[Y_i] \sigma[Y_j]}\,.
\label{eq:covariance_mat} 
\end{align}

Figure~\ref{fig:covmat} shows the correlation coefficients $r_{ij}$ corresponding to figure~\ref{fig:MR_uniM_cs2only}.
Panels (a) and (b) in figure~\ref{fig:covmat} show a sharp contrast particularly in $r_{23}$ and $r_{34}$, while the qualitative features of (c) are similar to those of (a).
One interpretation for this difference is as follows.
We see a general tendency for the inferred EoS to undergo a rapid stiffening at some threshold point of $\varepsilon$ possibly to support heavy NSs.
If the EoS is stiffer before the threshold, the EoS after the threshold tends to be softer, which is also favored to set the maximum mass and radius reasonably small.
In figure~\ref{fig:covmat}~(a) for the PREs/qLMXBs distribution, $r_{34}$ is negative reflecting that the stiffening threshold is found around $c_{s,n=3}^2$ in view of the blue data in figure~\ref{fig:MR_uniM_cs2only}.  In the same way, negative $r_{23}$ in figure~\ref{fig:covmat}~(b) for the HP/NICER/GW distribution is understood from the stiffening threshold at $c_{s,n=2}^2$ as shown by the green data in figure~\ref{fig:MR_uniM_cs2only}.

\section{Discussions} \label{sec:NN_property}
We discuss consistency checks of our results $\PrObs$ by manipulating our NN procedures.
To discuss uncertainty propagation, in section~\ref{sec:error_prop}, we perform test analysis with artificially rescaled $\PrObs$.
We then argue parametrization dependence for the discretization of $\varepsilon$ in section~\ref{sec:EoS_param} and distribution for sampling dataset and its prior dependence in section~\ref{sec:sample_dis}.
In section~\ref{sec:pc}, we extend the NN architecture with additional physical information of the central pressure $p_c$, which is useful to manifest the correlation between the NS properties and our results.  We also use the NN with $p_c$ to discuss noise dependence in the data augmentation in section~\ref{sec:noise}.  Finally, we comment on the direct evaluation of Jacobian as a possible alternative framework in section~\ref{sec:jac}.

\subsection{Test analysis with rescaled observational distributions} \label{sec:error_prop}

\begin{figure}
  \centering
  \includegraphics[width=0.9\textwidth]{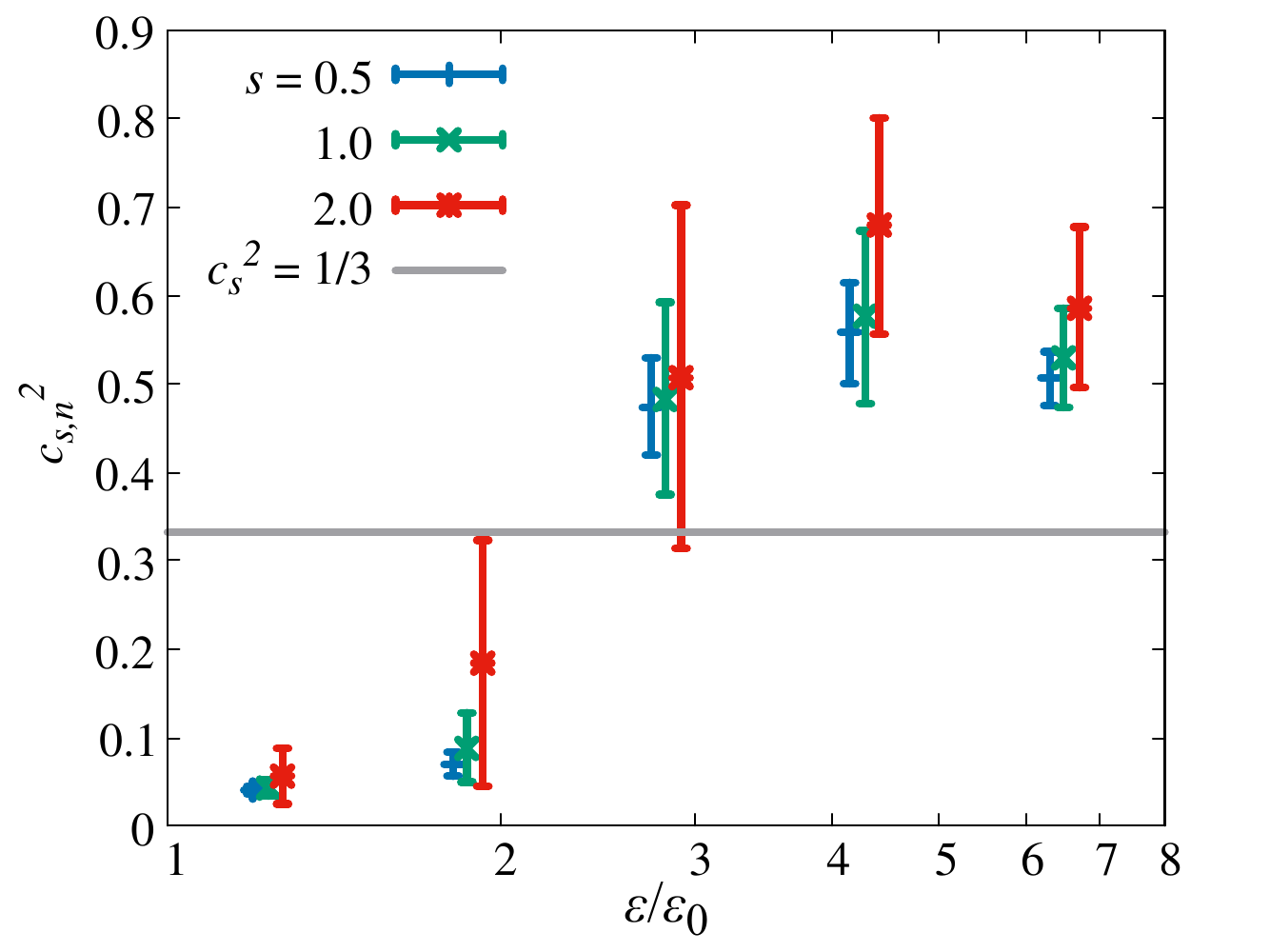}
  \caption{Speed of sound squared, $c_{s,n}^2$, resulting from rescaled $\PrObs$ in the MC integration using the full distribution with $\NSNumber=20$. The error bars show the standard deviation of $\bm{Y}$ over the modified distribution by replacing $\PrObs$ in eq.~\eqref{eq:outline.ansatz} with $\mathcal{P}_s$ defined in eq.~\eqref{eq:Psc}.}
\label{fig:resc_pex}
\end{figure}

We naturally expect that the uncertainty on the output side should be enhanced if the input has a larger uncertainty.  Here, we verify whether this expected behavior is properly realized in our NN method by modifying $\PrObs$ in the MC integration.
Specifically, we rescale $\PrObs(\bm{X})$ around its mode, $\hat{\bm{X}} = \arg \max_{\bm{X}} \PrObs(\bm{X})$ and introduce a scaling parameter $s \in \mathbb{R}_{>0}$.  We define a modified distribution as
\begin{align}
  \mathcal{P}_{s}(\bm{X})  \dif \bm{X} := s \PrObs(\hat{\bm{X}}+ s^{-1} \Delta \hat{\bm{X}} ) \dif \bm{X} \,, \qquad
  \Delta \hat{\bm{X}} := \bm{X} - \hat{\bm{X}}\,.
  \label{eq:Psc} 
\end{align}
Figure~\ref{fig:resc_pex} shows the results for $s=0.5$, $1.0$, and $2.0$.
We obtained the expected $s$-dependence, namely, the error bars become larger (smaller) with increasing (decreasing, respectively) $s$.
The uncertainty on the input side characterized by $\mathcal{P}_s$ reasonably propagates to the uncertainty on the output side quantified by the error bars in figure~\ref{fig:resc_pex}.
We also observe a trend of increasing mean value with increasing $s$.
This can be understood from the difference between the average and mode in $\PrObs$.

\subsection{Dependence on the number of segments in the EoS parametrization} \label{sec:EoS_param}

\begin{figure}
  \centering
  \includegraphics[width=0.9\textwidth]{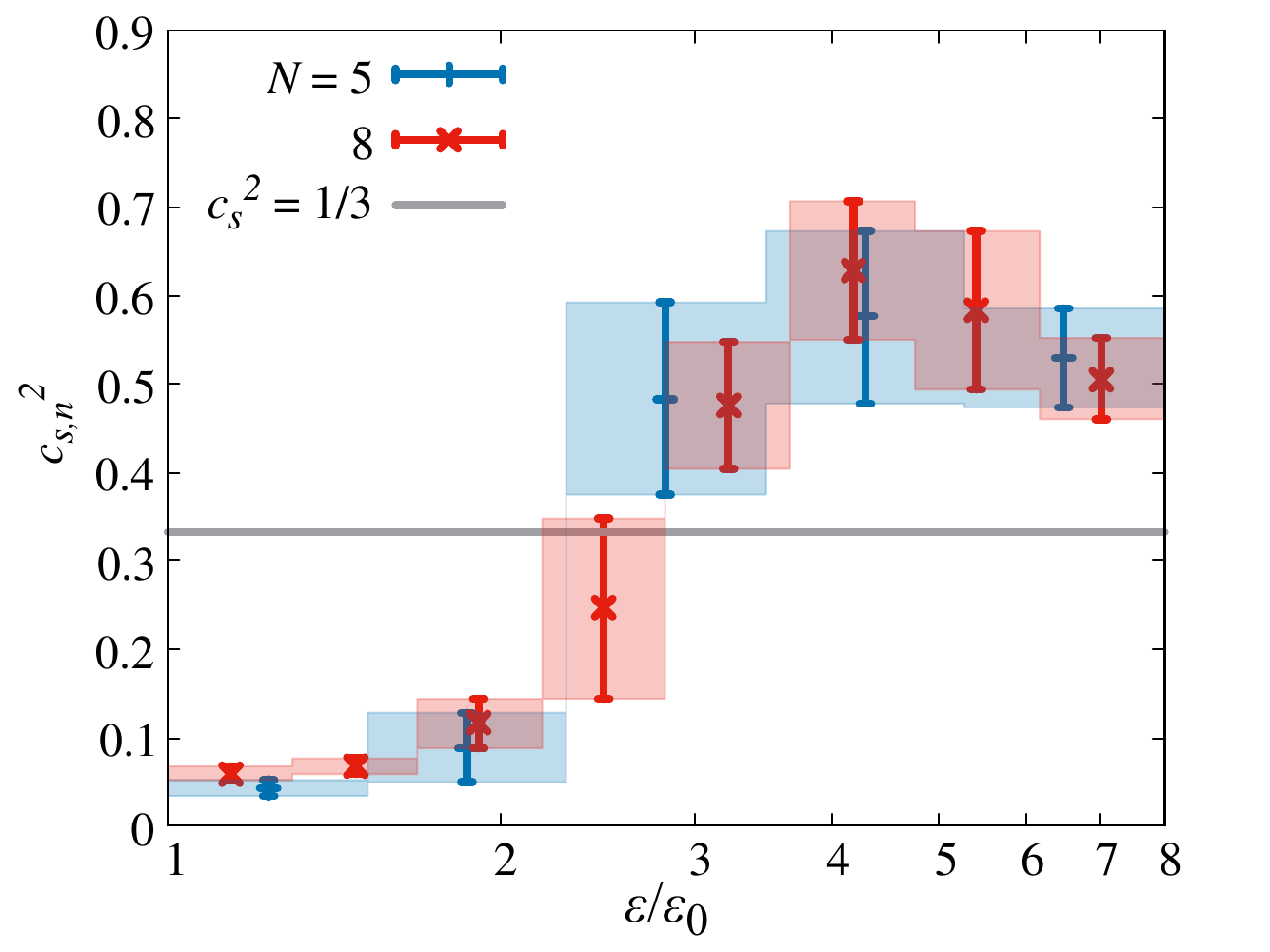}
  \caption{Comparison of $c_{s,n}^2$ estimated by $N$-segment polytropes with $N=5$ (blue) and $N=8$ (red). The error bars show the standard deviation of $\bm{Y}$ over $\PrEoS'$.}
  \label{fig:N_uniM_cs2only}
\end{figure}

As explained in section~\ref{sec:EoS}, we parametrize our EoS by dividing the range of density into $N$ segments
in the logarithmic scale, i.e., by a $N$-segment polytrope.
Throughout this work, we fix $N=5$ unless otherwise noted.
To check the bias on the specific choice of the EoS parametrization,
we can check the robustness of the results on different $N$.
We show a comparison between the cases with $N=5$ and $N=8$ in figure~\ref{fig:N_uniM_cs2only}.
Overall, the $N=8$ results are fairly consistent with the $N=5$ results, interpolating the $N=5$ points.
This implies that our NN architecture could be safely extended for a larger $N$
to capture more detailed features of EoS
when the available data is increased in both quantity and quality in the future.
The EoS rapidly rises around $\varepsilon/\varepsilon_0 \simeq 2$--$4$ for both $N=5$ and $N=8$, and we can safely conclude that the rapid increase in $c_s^2$ is physical and not caused by the EoS parametrization artifact.
The discussion on the optimal EoS parametrization will be more closely reported elsewhere.

\subsection{Training dataset sampling and prior dependence check}\label{sec:sample_dis}

\begin{figure}
  \centering
  \includegraphics[width=0.9\textwidth]{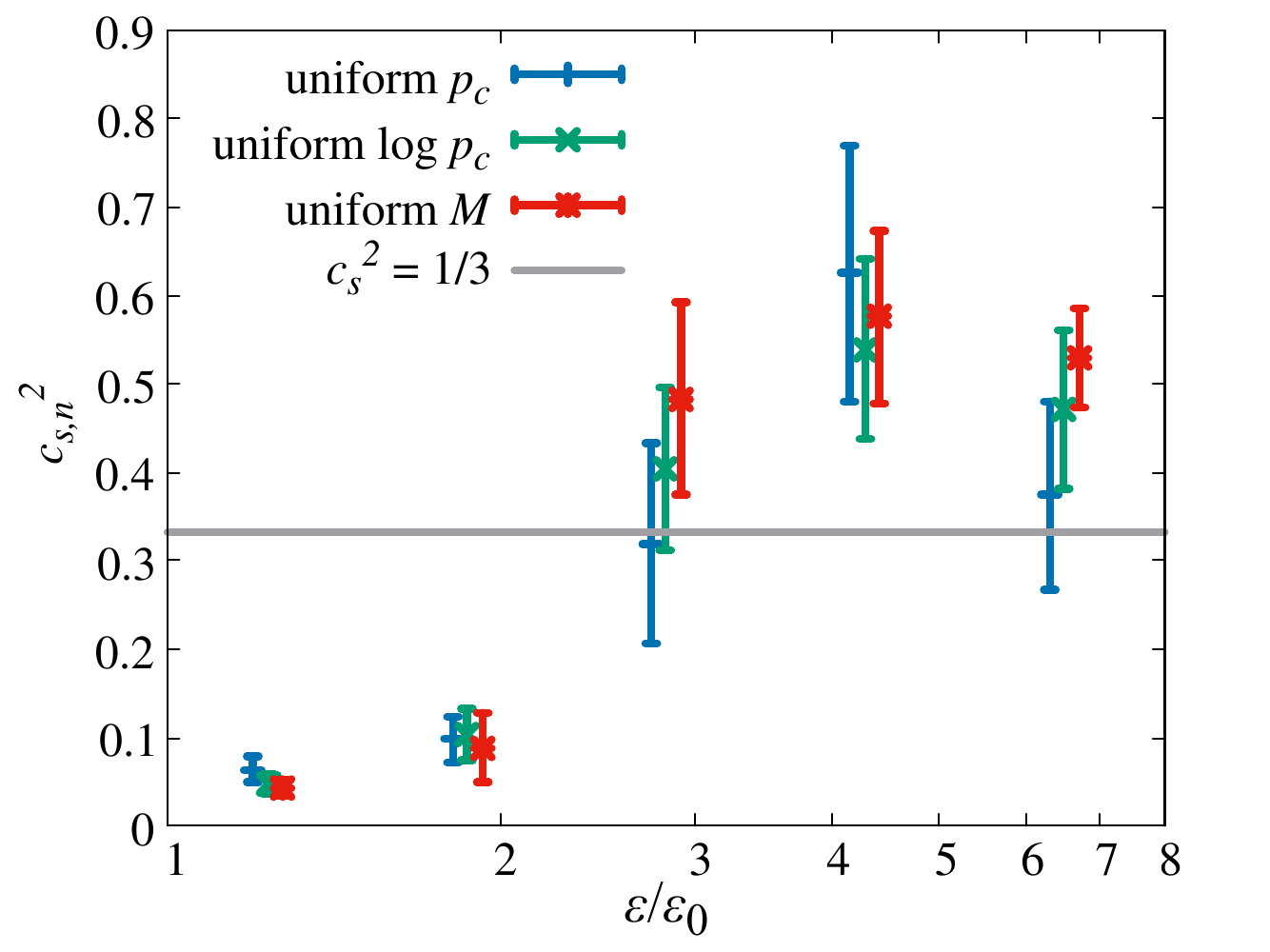}
  \caption{Dependence of $c_{s,n}^2$ on the distribution assumed for sampling $\NSNumber$ points from the $M$-$R$ curve.  For descriptions of the assumed distributions, see the text.
  The error bars show the standard deviation of $\bm{Y}$ over $\PrEoS'$.}
\label{fig:sample_dis}
\end{figure}

In the training data generation, as explained in step 3 in section~\ref{sec:method.dataset}, we sample $\NSNumber$ neutron stars from the uniform distribution in the range $[M_{\min}, M_{\max}]$, which is denoted by $\UniDis(M_{\min}, M_{\max})$.
We do not have \textit{a priori} knowledge about the distribution of NSs on the $M$-$R$ curve.
Our choice of the uniform distribution in mass is just a possible scenario.
This is a factor to comprise our prior assumption in the analysis.
We should avoid any biases by this distribution choice,
i.e., the results should not depend on our prior assumption if the EoS is well-constrained by the NS data.
To check this, we discuss the results with different choices for the distributions of sampled NS masses.

Here, we compare three different distributions.
The first choice is the uniform distribution, $\UniDis(M_{\min}, M_{\max})$, as adopted in the main analysis of this work.
Instead of mass, one can also specify a point on the $M$-$R$ curve through the NS central pressure $p_c$ as explained in eq.~\eqref{eq:pc}. Here, it would be reasonable to take $\UniDis(p_{c, {\min}}, p_{c, {\max}})$ as our second choice,
where $p_{c, \mathrm{min/max}}$ is fixed by the condition, $M(p_{c, \mathrm{min/max}})=M_{\mathrm{min/max}}$.
We also make a comparison with the log uniform distribution, namely, $\UniDis(\log p_{c, {\min}}, \log p_{c, {\max}})$ as the third choice.

In figure~\ref{fig:sample_dis}, we compare the results from different sampling distributions.
We observe that the results are consistent with each other within the error bars up to the fourth segment.
In the fifth (highest density) segment, $c_s^2$ depends on the sampling distributions relatively more strongly than the lower-density regions.
This implies that the fifth segment may not be sufficiently constrained by the current analysis.  One may think that the available data has no sensitivity to such high-density regions, but this is not necessarily the case as we demonstrate in the next subsection.

\subsection{Extension with the NS central pressure} \label{sec:pc}

\begin{figure}
  \centering
  \includegraphics[width=0.9\textwidth]{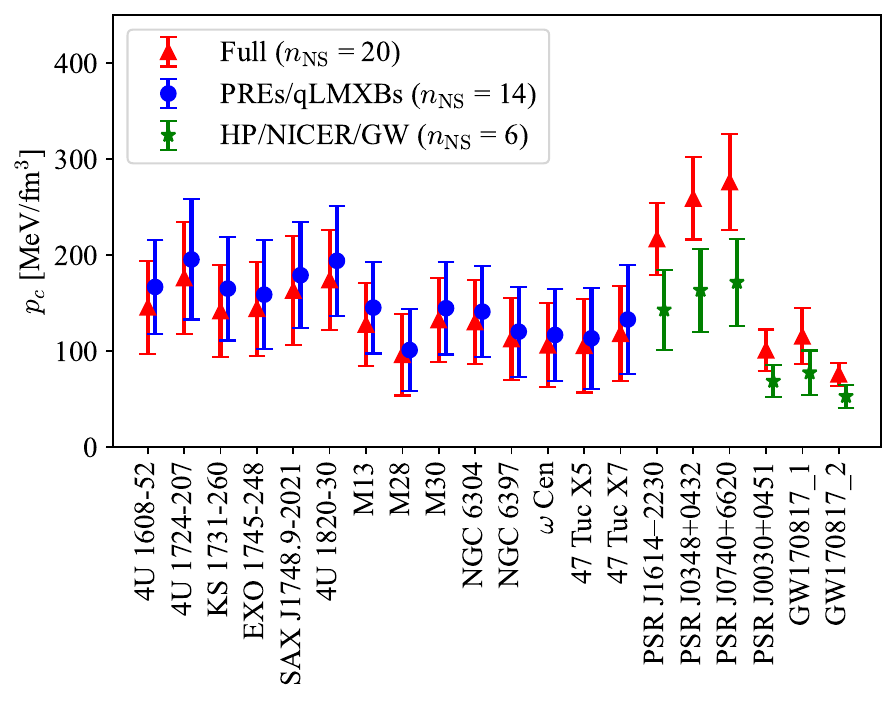}
  \caption{Central pressure of the NSs from the NN extended with $p_c$.  The results inferred from the full, PREs/qLMXBs, and HP/NICER/GW distributions are shown by the red, blue, and green bars, respectively, which represent the standard deviation of $\bm{Y}$ over $\PrEoS'$.}
  \label{fig:extend_pc}
\end{figure}

As discussed in section~\ref{sec:outline}, the uncertainty in the EoS chiefly comes from two factors,
i.e., the finiteness of the observed NS number and the limitation of observational constraints on the observed NS properties.
Some randomness is inevitable in determining both the true NS properties and the observation process.
Although our primary interest is the true EoS inferred in principle from $\{(M_*^{(i)}, R_*^{(i)})\}_{i=1}^{\NSNumber}$, the other NS properties, if any, should also be physical information, which would be precisely determined by the data if the observation had unlimited resolution.
To have an insight into the randomness involved in the EoS determination, it is useful to check how the other NS properties are estimated and how they are correlated to the estimated EoS within our method.
Here, for the demonstration purpose, we choose a single parameter, i.e., the central pressure $p_c$ to represent other NS properties on top of the EoS parameters.

Here, we extend our NN model in section~\ref{sec:Arch} to include the NS central pressures, $\{p_c^{(i)}\}_{i=1}^{\NSNumber}$, in the output layer.  Specifically, the output data is
\begin{align}
 \bm{Y} &:=  \{ c_{s,n=1}^{2}, \dots, c_{s,n=5}^{2},\, p_c^{(1)}, \dots, p_c^{(\NSNumber)} \}\,, \label{eq:Oout_pc}
\end{align}
where $p^{(i)}_c$ is given in unit of $\text{MeV}/\text{fm}^3$ and determined by $(M_*^{(i)}, R_*^{(i)})$ in step 3 in section~\ref{sec:method.dataset}.
We optimize this model in the same way as previously but with the dataset extended with $\{p_c^{(i)}\}_{i=1}^{\NSNumber}$.

Figure~\ref{fig:extend_pc} shows the results for the full distribution ($\NSNumber=20$ by the red bars), the PREs/qLMXBs distribution ($\NSNumber=14$ by the blue bars), and the HP/NICER/GW distribution ($\NSNumber=6$ by the green bars).
The mean values and errors are evaluated by eq.~(\ref{eq:mean_error}).
In the full results, $p_c$'s for the HP/NICER/GW NSs indicated by the rightmost six red bars in figure~\ref{fig:extend_pc} are larger than $p_c$'s from the HP/NICER/GW distribution only indicated by the green bars in figure~\ref{fig:extend_pc}.
This enhancement of $p_c$ for the HP/NICER/GW NSs in the full case is caused by the inclusion of the PREs/qLMXBs NSs.
This also explains the difference in the $c_s^2$ behavior inferred from the full and the partial HP/NICER/GW distributions shown in figure~\ref{fig:MR_uniM_cs2only}.

Figure~\ref{fig:extend_pc} also tells us the highest density range that each NS can reach.
From figure~\ref{fig:eos_result}, one can read out the central energy density $\varepsilon_c$ corresponding to $p_c$ of the red bars in figure~\ref{fig:extend_pc}.
The largest $p_c$ is realized for J0740+6620, and the corresponding energy density turns out to be $\varepsilon_c \simeq 5.6 \varepsilon_0$, which is larger than the boundary of the fifth segment at $\varepsilon \simeq 5.2 \varepsilon_0$.
Therefore, on the one hand, in the current analysis with the full distribution, we can say that the observational data constrains the EoS up to the fifth segment.
On the other hand, in the previous analysis, we only used the PREs/qLMXBs NS data.
The central density $\varepsilon_c$ corresponding to the largest $p_c$ read from the maximum value of the blue bars in figure~\ref{fig:extend_pc} is smaller than the boundary of the fifth segment.
Therefore, we confirm that the fifth segment of the EoS in our previous works~\cite{Fujimoto:2019hxv, Fujimoto:2021zas} was not well constrained by the NS data, which is consistent with our previous discussion in ref.~\cite{Fujimoto:2021zas}.

A similar conclusion can be drawn for the EoS inferred from the HP/NICER/GW NS data only.
The central density $\varepsilon_c$ corresponding to the maximum $p_c$ of the green bars in figure~\ref{fig:extend_pc} is smaller than the boundary of the fifth segment.  Thus, the fifth segment of the EoS may not be constrained.
This explains why the green error bar in the fifth segment in figure~\ref{fig:MR_uniM_cs2only} is small around $c_{s,n}^2\simeq 0.5$.

\subsection{Noise dependence in data augmentation} \label{sec:noise}

\begin{figure}
  \centering
    {\Large (a)}\\
    \includegraphics[clip, width=0.75\textwidth]{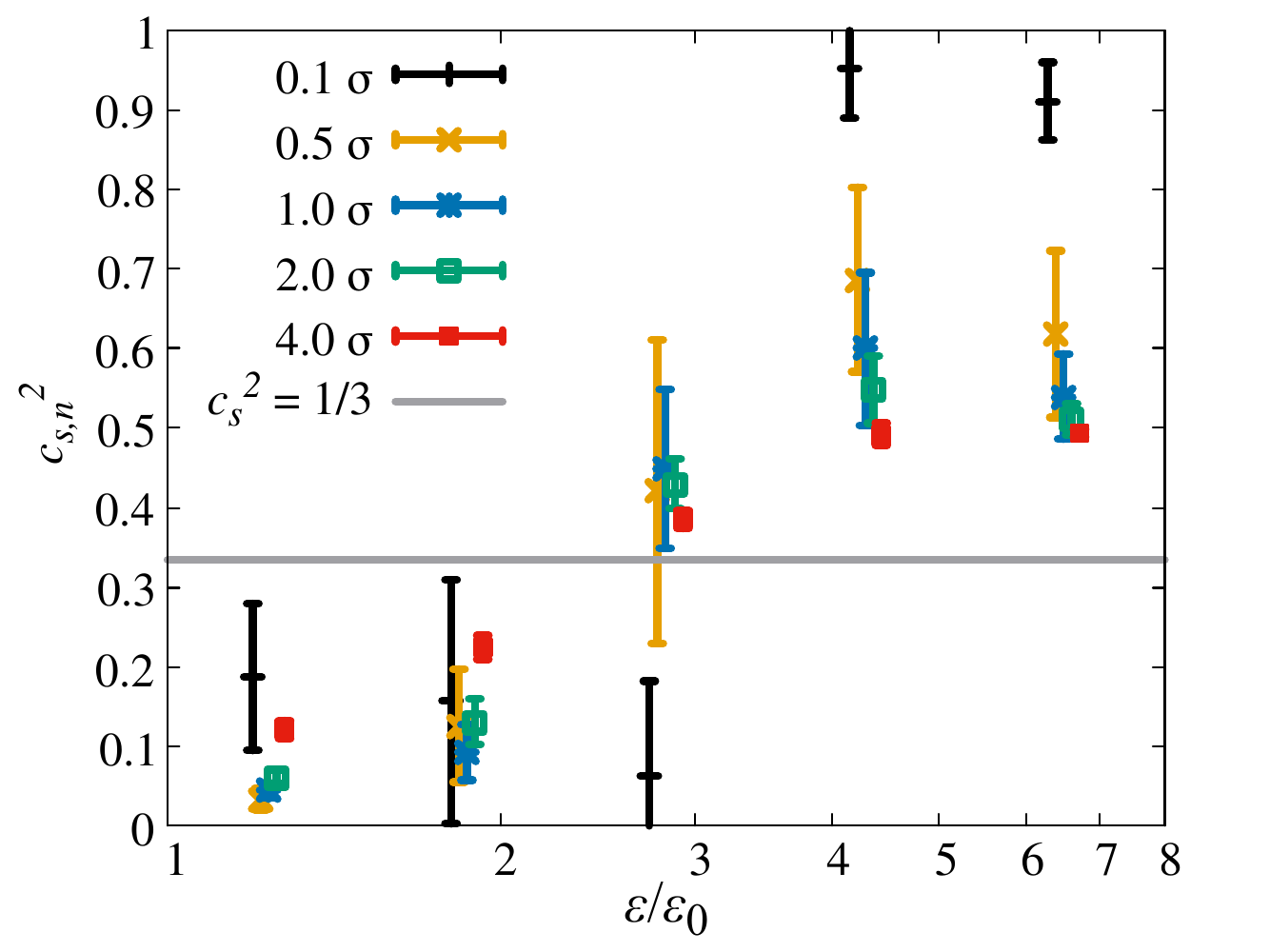}
    \vspace{1em}\\
    {\Large (b)}\\
    \includegraphics[clip, width=0.7\textwidth]{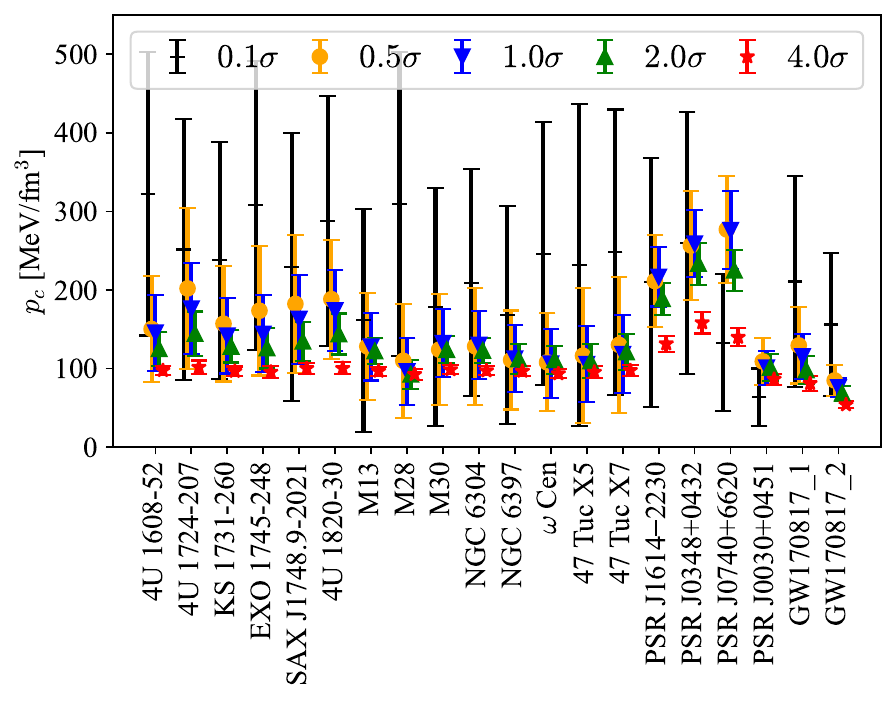}
    \caption{Noise dependence of (a) $c_{s,n}^2$ and (b) $p_c$ for the full distribution with $\NSNumber=20$.
    The noises are generated by the Gaussian distribution with the standard deviation $s \sigma$ modified by a parameter $s$.  The plotted results are for $s=0.1$ (black), $0.5$ (orange), $1.0$ (blue), $2.0$ (green), and $4.0$ (red). 
    The error bars show the standard deviation of $\bm{Y}$ over $\PrEoS'$.}
    \label{fig:compare_sigma}
\end{figure}

We discuss noise dependence in data augmentation using the NN model extended with $p_c$.
The inclusion of $p_c$ is useful to clarify the effect of the data augmentation, which is important in the training data generation.
As explained in step 4 in section~\ref{sec:method.dataset}, we randomly sample $\NSNumber$ observational points from the Gaussian distribution.  A concern with this method is that there is no inherent way to fix the noise magnitude.
Below, we quantitatively study the effect of changing the noise magnitude on the inferred results.

We vary the noise size by changing the training dataset with $s\in\mathbb{R}_{>0}$ as
\begin{align}
  ({M}^{(i)},R^{(i)}) \;\sim\; \mathcal{N} (M_*^{(i)}, ( s \sigma_{M}^{(i)} )^2 ) \times \mathcal{N} (R_*^{(i)}, ( s \sigma_{R}^{(i)} )^2 )\,.
  \label{eq:Noise_s} 
\end{align}
We show the results for $s=0.1$, $0.5$, $1.0$, $2.0$, and $4.0$ in figure~\ref{fig:compare_sigma}.
In panel (a) of figure~\ref{fig:compare_sigma}, we see that the error bars are larger for smaller $s$.
The behavior of the sound velocity also does not appear to be reasonable when $s$ is too small $\sim 0.1\sigma$;
the sound velocity is mostly vanishing in the third segment,
while it is mostly the light velocity in the fourth and fifth segments.
This is because the NN trained with small $s$
does not properly learn the existence of the observational errors
and expects the points $(M^{(i)}, R^{(i)})$ in the input to align on a single curve.
The NN with small $s$ cannot properly handle the unaligned input $\bm{X}$ generated by $\PrObs$
and starts to produce biased results.

The results do not seem sensible either in the opposite case of large $s \sim 4$;
the error bars are unnaturally small.
This is because the NN learns that the input is not trustable when it is trained by large $s$
so tends to predict a constant output without using the input.
In other words, the NN trained by large $s$ outputs
a constant prediction mainly determined by the prior distribution,
to which the structures of the input distribution $\PrObs$ are not reflected.
In particular, this loss of the input distribution can be observed
in the results of the central pressures, $p_c$, as shown in the panel~(b) of figure~\ref{fig:compare_sigma}.
The results of $p_c$ for $s=4.0$ are located around $p_c \sim 100\;{\rm MeV/fm^3}$ only.
This implies that data augmentation with too large noise unreasonably undervalues the meaningful information of NS features in the observation data.
This noise dependence could be relevant to other prior dependence.
We will further investigate this issue in our future work.

\subsection{Alternative framework: direct evaluation of Jacobian} \label{sec:jac}

Finally, we mention a possible alternative framework.
In section~\ref{sec.outline.strategy}, we constructed the inverse mapping of the TOV equation $\MapInv$ in eq.~\eqref{eq:outline.point-regression} to obtain the most likely EoS\@.
We already know the forward mapping $\MapTOV: \bm{Y} \mapsto \bm{X}$ by solving the TOV equation,
so that we can construct the \emph{pullback} of $\PrObs$ by $\MapTOV$ to obtain $\PrEoS$, i.e.,
\begin{align}
  \PrEoS(\bm{Y}) \dif\bm{Y}
  = (\MapTOV^* \dif\PrObs)(\bm{Y})
  = \PrObs(\MapTOV(\bm{Y})) \biggl|\frac{\dif\MapTOV(\bm{Y})}{\dif\bm{Y}}\biggr| \dif\bm{Y}\,.
  \label{eq:outline.pullback}
\end{align}
In reality, the above formula should be generalized to take account of randomness in the observation uncertainty:
\begin{align}
  \PrEoS(\bm{Y}) \dif\bm{Y}
  \sim \int \dif\mathcal{P}(\omega) \PrObs(\MapTOV(\bm{Y}, \omega)) \biggl|\frac{\dif\MapTOV(\bm{Y}, \omega)}{\dif\bm{Y}}\biggr| \dif\bm{Y}\,,
\end{align}
where $\omega$ symbolically denotes the random variables involved in sampling $M$ with observational errors.  In this framework, the inversion is incorporated in the pullback, so an explicit construction of the inverse mapping is unnecessary.
Historically speaking, this pullback method, in which one must explicitly compute the Jacobian, was the first approach taken in the realm of Bayesian inference~\cite{Ozel:2009da, Ozel:2010fw}.

Indeed, if the inversion were exact (i.e., $\MapInv \circ \MapTOV = \mathrm{id}_\mathrm{EoS}$ and $\MapTOV \circ \MapInv = \mathrm{id}_{MR}$), the above definition by the pullback exactly reproduces eq.~\eqref{eq:outline.ansatz}.  In practice, however, the evaluation of the Jacobian is a nontrivial numerical task, and our method presented in this work is computationally more straightforward.

\section{Conclusions} \label{sec:conclusions}

We developed a new method to properly incorporate the shape of the joint probability distribution of NS masses and radii, which we denoted $\PrObs$, by performing the MC integration.
The new method naturally incorporates the uncertainty quantification in the final output from the NN, which had been more like a black box in our previous method.
The results in the present work imply that the previous evaluation of the output error by bootstrap aggregating gives an estimate consistent with the current method.

The chief result of this work is presented in figure~\ref{fig:MR_uniM_cs2only}.
The inclusion of new NS data upon the older ones, particularly the $M$-$R$ data from the NICER and GW170817 event, led to a sharp change in the stiffness of the EoS as displayed in figure~\ref{fig:eos_result} (see also the error bars marked as ``Full ($\NSNumber = 20$)'' in figure~\ref{fig:MR_uniM_cs2only}).
This is quantified by the weaker correlation between the second and third segments of the piecewise polytrope shown in panel (c) of figure~\ref{fig:covmat} compared to panel (a).
Our previous analysis of observational data suggested a peak structure in the speed of sound squared, $c_s^2$, as a function of energy density, and the current analysis including new data strengthens this claim, though the statistical significance is not yet achieved with the current accuracy of the NS data.

It is an important question up to which density the observational data can in principle constrain.
In the present study, 
the decrease of the sound speed at the largest energy density range, in particular, is certainly a feature under the constraint.
This contrasts with our previous work.
This decrease in the previous work may have been attributed to the prior assumptions due to the lack of constraining data.
In the new analysis, we substantiated that the $c_s^2$ behavior in this high-density region is not a mere reflection of the prior assumptions but is also constrained by the new data:
First, we changed the uncertainty in the input data and verified that the input uncertainty is properly propagated to the output uncertainty as one can see in figure~\ref{fig:resc_pex}.
This fact implies that the last segment is also constrained by the new data.
Indeed, as shown in figure~\ref{fig:extend_pc}, the pulsar J0740+6620 is predicted to have the central pressure $p_c \approx 300~\text{MeV/fm}^3$ in our analysis, which roughly corresponds to the energy density in the last segment in our parametrization.

Furthermore, we examined three artificially adjusted prior assumptions to see the effects on the EoS\@.
We demonstrated that the prior assumptions do not affect the results within the uncertainties.
First, we changed the number of segments in the piecewise polytrope.
We verified in figure~\ref{fig:N_uniM_cs2only} that the results are insensitive to the number of segments within the uncertainty.
Second, for the input to the NN, we changed the distributions for sampling $\NSNumber$ points along the $M$-$R$ curve.
This change may lead to substantial prior dependence, but as seen in figure~\ref{fig:sample_dis}, the results are still consistent with each other within the uncertainties.
Finally, we changed the magnitude of the noise used in the data augmentation.
In figure~\ref{fig:compare_sigma}, we found that too small noise leads to physically unacceptable results, but too large noise also leads to unconstrained results with underestimated output uncertainty.
The current choice of noise comparable to the observational error ($1.0\sigma$) turns out to be the most reasonable choice.
Related to this, we also note that our new method requires a smaller size of the training dataset because this MC integration can lessen the magnitude of the data augmentation necessary in the uncertainty analysis.
A mathematically founded method to optimize the data augmentation procedure to properly incorporate the effect of uncertainty in the NN is yet to be explored.
More detailed investigations of these prior effects will be presented elsewhere.

\acknowledgments
We thank Len~Brandes, Tetsuo~Hatsuda, and Shuhei~Minato for useful discussions.
The work of Y.F.\ is supported by the Japan Society for the Promotion of Science (JSPS) through the Overseas Research Fellowship and by the INT's U.S.\ DOE Grant No.~DE-FG02-00ER41132.
This work was supported by JSPS KAKENHI Grant Nos.\ 22H01216 (K.~F), 22H05118 (K.~F and S.~K), and 23K13102 (K.M.).

\bibliographystyle{JHEP}
\bibliography{nn}

\end{document}